\documentclass[aps,onecolumn,superscriptaddress,prl,floatfix,british,10pt]{revtex4-2}
\usepackage[dvips]{graphicx} % for figures

\usepackage{amsfonts}
\usepackage{amssymb}
\usepackage{amscd}
\usepackage{amsmath}    % need for subequations
\usepackage{enumerate}
\usepackage{epsfig}
\usepackage{subfigure}
\usepackage{subfloat}
\usepackage{xcolor}
\usepackage{amsthm}
\usepackage{physics}
\usepackage{diagbox, multirow}
\usepackage[most]{tcolorbox}
\usepackage[]{qcircuit}
\usepackage{tabularx}
\usepackage{float}
\usepackage{appendix}
\usepackage{makecell}
\usepackage{algorithm}
\usepackage{algorithmicx}
\usepackage{algpseudocode}
\usepackage{dsfont}
\usepackage[colorlinks, linkcolor=red, anchorcolor=blue, citecolor=green]{hyperref}
\usepackage{MnSymbol}
\newtcolorbox[auto counter]{mybox}[2]{
	label=#1,
	enhanced,
	title=Box \thetcbcounter: #2,
	fonttitle=\bfseries,
	colback=blue!5!white,
	colframe=blue!75!black
}

\begin{document}
\title{Calibrating Quantum Gates up to 52 Qubits in a Superconducting Processor}
\author{Daojin Fan}\thanks{These authors contributed equally to this work.}
\affiliation{\adda}
\affiliation{\addb}
\affiliation{\addc}
\author{Guoding Liu}\thanks{These authors contributed equally to this work.}
\affiliation{\addd}
\author{Shaowei Li}\thanks{These authors contributed equally to this work.}
\affiliation{\adda}
\affiliation{\addb}
\affiliation{\addc}
\author{Ming Gong}
\affiliation{\adda}
\affiliation{\addb}
\affiliation{\addc}
\affiliation{\addf}
\author{Dachao Wu}
\affiliation{\adda}
\affiliation{\addb}
\affiliation{\addc}
\author{Yiming Zhang}
\affiliation{\adda}
\affiliation{\addb}
\affiliation{\addc}
\author{Chen Zha}
\affiliation{\adda}
\affiliation{\addb}
\affiliation{\addc}
\author{Fusheng Chen}
\affiliation{\adda}
\affiliation{\addb}
\affiliation{\addc}
\author{Sirui Cao}
\affiliation{\adda}
\affiliation{\addb}
\affiliation{\addc}
\author{Yangsen Ye}
\affiliation{\adda}
\affiliation{\addb}
\affiliation{\addc}
\author{Qingling Zhu}
\affiliation{\adda}
\affiliation{\addb}
\affiliation{\addc}
\author{Chong Ying}
\affiliation{\adda}
\affiliation{\addb}
\affiliation{\addc}
\author{Shaojun Guo}
\affiliation{\adda}
\affiliation{\addb}
\affiliation{\addc}
\author{Haoran Qian}
\affiliation{\adda}
\affiliation{\addb}
\affiliation{\addc}
\author{Yulin Wu} 
\affiliation{\adda}
\affiliation{\addb}
\affiliation{\addc}
\author{Hui Deng} 
\affiliation{\adda}
\affiliation{\addb}
\affiliation{\addc}
\affiliation{\addf}
\author{Gang Wu}
\affiliation{\adda}
\affiliation{\addb}
\affiliation{\addc}
\affiliation{\addf}
\affiliation{\adde}
\author{Cheng-Zhi Peng}
\affiliation{\adda}
\affiliation{\addb}
\affiliation{\addc}
\affiliation{\addf}
\author{Xiongfeng Ma}
\thanks{xma@tsinghua.edu.cn}
\affiliation{\addd}
\author{Xiaobo Zhu}
\thanks{xbzhu16@ustc.edu.cn}
\affiliation{\adda}
\affiliation{\addb}
\affiliation{\addc}
\affiliation{\addf}
\author{Jian-Wei Pan}
\thanks{pan@ustc.edu.cn}
\affiliation{\adda}
\affiliation{\addb}
\affiliation{\addc}
\affiliation{\addf}

\newcommand{\adda}{Hefei National Research Center for Physical Sciences at the Microscale and School of Physical Sciences, University of Science and Technology of China, Hefei 230026, China}
\newcommand{\addb}{Shanghai Research Center for Quantum Science and CAS Center for Excellence in Quantum Information and Quantum Physics, University of Science and Technology of China, Shanghai 201315, China}
\newcommand{\addc}{Hefei National Laboratory, University of Science and Technology of China, Hefei 230088, China}
\newcommand{\addd}{Center for Quantum Information, Institute for Interdisciplinary Information Sciences, Tsinghua University, Beijing, 100084 China}
\newcommand{\adde}{University of Science and Technology of China, Shanghai Research Institute, Shanghai 201315, China}
\newcommand{\addf}{Jinan Institute of Quantum Technology and Hefei National Laboratory Jinan Branch, Jinan 250101,China}

\begin{abstract}
Benchmarking large-scale quantum gates, typically involving multiple native two-qubit and single-qubit gates, is crucial in quantum computing. Global fidelity, encompassing information about inter-gate correlations, offers a comprehensive metric for evaluating and optimizing gate performance, unlike the fidelities of individual local native gates. In this work, utilizing the character-average benchmarking protocol implementable in a shallow circuit, we successfully benchmark gate fidelities up to 52 qubits.
% , marking the largest gate benchmarking experiment to date. 
Notably, we achieved a fidelity of $63.09\%\pm0.23\%$ for a 44-qubit parallel CZ gate. Utilizing the global fidelity of the parallel CZ gate, we explore the correlations among local CZ gates by introducing an inter-gate correlation metric, enabling one to simultaneously quantify crosstalk error when benchmarking gate fidelity. Finally, we apply our methods in gate optimization. By leveraging global fidelity for optimization, we enhance the fidelity of a 6-qubit parallel CZ gate from 87.65\% to 92.04\% and decrease the gate correlation from 3.53\% to 3.22\%, compared to local gate fidelity-based optimization. The experimental results align well with our established composite noise model, incorporating depolarizing and $ZZ$-coupling noises, and provide valuable insight into further study and mitigation of correlated noise.
\end{abstract}

\maketitle

\section{Introduction}
Realistic quantum computers suffer from noise, hindering themselves from demonstrating an advantage over their classical counterpart~\cite{aharonov1996limitations,Alexander2016entropy,Daniel2021Limitations,Aharonov2023Noisy,yan2023limitations}. Quantum error correction is the key to reducing noise~\cite{Egan2021corrected,Gong2021correcting,Postler2022tolerant,Zhao2022Correcting,Acharya2023Suppressing}, yet its implementation hinges on the availability of high-fidelity quantum gates and weak correlations among different gate operations~\cite{Kitaev1997computations,aharonov1997fault,Knill1998thresholds}. Thus, diagnosing the noise level and assessing the interaction within the quantum circuits is essential to improving their performance and, hence, realizing fault-tolerant quantum computing.
Within a step of quantum circuits, single-qubit and two-qubit gates are parallelly implemented to reduce the circuit depth. Due to the potential crosstalk and residual coupling between the qubits~\cite{barends2014logic}, the local gate fidelities may be insufficient to give an overall evaluation of the global gate. In contrast, global gate fidelity contains information about correlations among different local gates, providing an overall performance metric.

As quantum hardware advances and platforms grow, large-scale quantum gate calibration becomes increasingly crucial. Researchers have made great efforts to develop practical benchmarking methods~\cite{Chuang1997tomo,Flammia2011prlDirectFidelity,Emerson2005Scalable} and enlarge the gate calibration size. Currently, the most frequently used method to evaluate the quantum gate fidelity is randomized benchmarking (RB)~\cite{Emerson2007Characterization, Knill2008RB, Emerson2011prlRB, Emerson2012praRB,Magesan2012interleavedRB}, which enjoys the advantage of low complexity and robustness to state preparation and measurement errors. Nonetheless, the original RB protocol~\cite{Emerson2011prlRB} is only realized up to three qubits~\cite{McKay2019Three} owing to the compiling problem of global Clifford gates. Several RB variants~\cite{Proctor2019DirectRB,Erhard2019cycleRB,hines2022demonstrating} were proposed circumventing compiling issues and significantly advanced the benchmarking size to 10 qubits for an individual gate via cycle benchmarking~\cite{Erhard2019cycleRB} and 27 qubits for a gate set via mirror RB~\cite{hines2022demonstrating}.

Nonetheless, a gap exists between these achievements and the scale of state-of-the-art quantum computers. Currently, quantum computers have been realized with tens of or even hundreds of qubits in superconducting~\cite{ZHU2022advantage,morvan2024phase,mckay2023benchmarking}, ion trap~\cite{Moses2023Ion,Joshi2023large}, and neutral atom systems~\cite{Bluvstein2024atom}. Calibrating larger gates is demanding for assessing and enhancing the performance of current quantum computers. Meanwhile, a noticeable challenge to achieving this goal is the low fidelity often associated with large-scale gates. Many RB protocols require repeated execution of the target gate. However, noise can prevent the gate from being repeated without significant signal loss, which can render the protocol ineffective. Resolving this issue requires enhancing benchmarking protocols to ensure their effectiveness in short-depth circuits and achieving the realization of higher-performance quantum gates.

In this work, to achieve large-scale gate calibration, we utilize the character-average benchmarking (CAB)~\cite{Zhang2023cab} protocol. This method can evaluate the fidelity robust to state preparation and measurement errors for an individual Clifford gate up to a local unitary transformation and be scalable with respect to the system size. Compared to cycle benchmarking~\cite{Erhard2019cycleRB} that also aims to evaluate the individual gate fidelity, CAB requires shorter circuit depth and can tolerate higher gate errors and benchmarking larger gates. Additionally, CAB exhibits smaller statistical fluctuations than cycle benchmarking, with more details shown in the Supplemental Material~\cite{supplementary}. Meanwhile, in experiments, we improve gate fidelity and reduce a significant portion of gate control errors via qubit frequency adjustment and pre-calibration of gate parameters, as elaborated in Methods. The high fidelity and nearly depolarizing noise of the quantum gates are crucial to our success in realizing large-scale CAB experiments.

Here, we consider two important types of gates for experimental benchmarking. One is a fully connected gate composed of two layers of CZ gates and two layers of local Clifford gates. This gate is a part of the brickwise architecture circuit with an efficient realization scheme on current devices and hence is favored in variational quantum algorithms~\cite{Cerezo2021VQA}. It also plays an essential role in other quantum information processing tasks like simulating a nearest-neighbor interacting Hamiltonian evolution. We benchmark such gates up to 46 qubits and get a fidelity of $17.42\%\pm 0.45\%$ dressed with local twirling gates.

We also benchmark the parallel CZ gate, composed of parallelly implemented local CZ gates, which excels in generating entanglement across multiple parties and is essential in preparing graph states and executing numerous quantum algorithms~\cite{Raussendorf2003mbqc,Hein2004graph}. We benchmark such gates from 4 to 44 qubits. The average fidelity of a single local CZ gate is about $98\%$ and does not decrease with the qubit number increase.

With the global fidelity of the parallel CZ gate, we characterize the correlation among the local CZ gates. This procedure can be done simultaneously when benchmarking global fidelity without extra experiments. The correlation data provides the interaction information within the circuit and helps to evaluate and optimize the gate performance. The correlation of the 44-qubit parallel CZ gate turns out to be weak and constantly positive. In contrast, the results of the 52-qubit parallel CZ gate present some negative values. The experimental results are well explained by our established composite noise model, which incorporates local depolarizing and $ZZ$-coupling noises. The correlation magnitude positively depends on the coupling strength. Interestingly, the correlation sign between two gates varies in two cases: staying positive in two-gate coupling yet turning negative in three-gate coupling when one gate strongly couples with a third one. The correlation results help identify coupling gates and advance the study of correlated noise.

Since an important application of fidelity benchmarking is gate optimization, we demonstrate optimization experiments for parallel CZ gates and compare outcomes when setting the target function as the parallel CZ gate fidelity and the individual local CZ gate fidelities. We observe better results of the former. This result validates the effectiveness of CAB and suggests that global fidelity is more effective in optimizing quantum circuit performance, originating from containing more correlation information. The optimization result is consistent with the $ZZ$-coupling noise model. When three gates couple, improving the fidelity of one local CZ gate may reduce the fidelities of others. This antagonistic relationship demonstrates the limitations of local fidelities.

\section{Results}
\subsection{Preliminary}
% \textit{Preliminary.}---
Let us start with briefly revisiting the concept of gate fidelity. Any noisy quantum gate $\widetilde{U}$ can be treated as a composite of its noise channel $\Lambda$, and the ideal gate $U$, expressed as $\widetilde{U} = U\circ \Lambda$. In this work, the fidelity of gate $U$ refers to the process fidelity of $\Lambda$, defined by,
\begin{equation}\label{eq:processfidelity}
F(\Lambda) = 2^{-3n} \sum_{P\in \mathsf{P}_n} \tr(P\Lambda(P)),
\end{equation}
where $n$ is the number of qubits, and $\mathsf{P}_n = \{\mathbb{I}, X, Y, Z\}^{\otimes n}$ represents the $n$-qubit Pauli group. This process fidelity is a linear function of the average fidelity, a metric commonly employed in quantum gate benchmarking studies. The summation in Eq.~\eqref{eq:processfidelity}, initially spanning $4^n$ terms, can be effectively restructured into $2^n$ terms as follows:
\begin{equation}
\begin{split}
F(\Lambda) &= 2^{-2n} \sum_{w\in \{0, 1\}^{\otimes n}} 3^{\abs{w}}\lambda_w;\\
\lambda_w &= \frac{2^{-n} \sum_{\mathrm{pt}(P) = w} \tr(P\Lambda(P))}{3^{\abs{w}}},
\end{split}
\end{equation}
where $\mathrm{pt}(P)$ denotes a bitstring that takes 0 on bit $i$ if $P$ acts as the identity $\mathbb{I}$ on qubit $i$ and takes 1 if $P$ acts non-trivially, and $\abs{w}$ signifies the weight of the bitstring $w$. The quantity $\lambda_w$ is termed the weighted quality parameter of $\Lambda$ with weight $2^{-2n}3^{\abs{w}}$, and below, we concisely refer to it as the quality parameter of $\Lambda$.

\begin{figure}[!htbp]
\centering \resizebox{17cm}{!}{\includegraphics{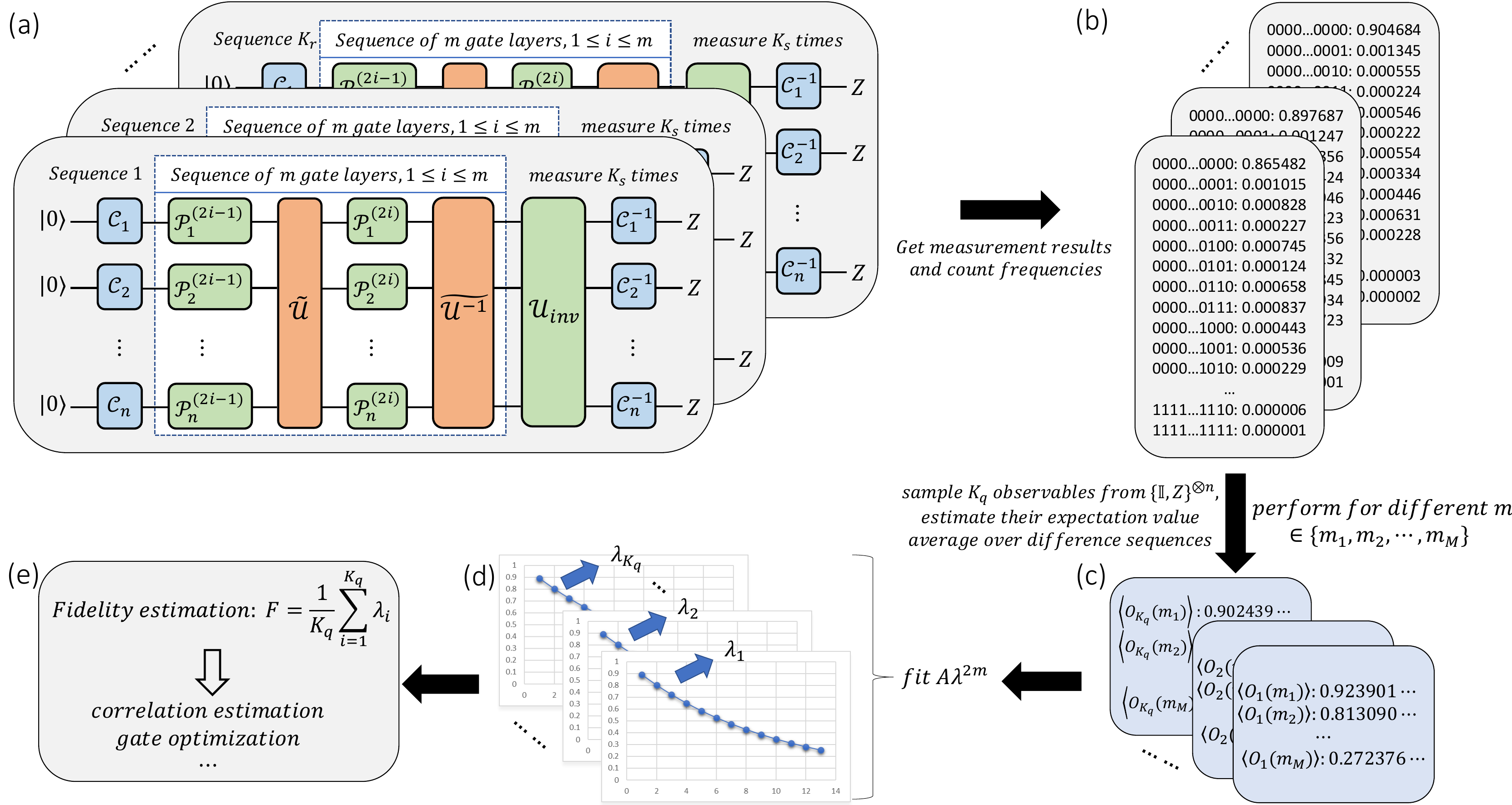}}
\caption{\textbf{Illustration of the CAB procedure for assessing the fidelity of an $n$-qubit gate, $U$.} (a) shows the circuit, beginning with the preparation of the state $\ket{0}^{\otimes n}$, followed by a random local Clifford gate $\bigotimes_{i=1}^n C_i$. Subsequently, $2m$ layers of random Pauli gates $\bigotimes_{j=1}^n P_j^{(i)}$ are interleaved with alternate sequences of $U$ and $U^{-1}$. The inverse gate $U_{inv} = (\Pi_{i=1}^m (U^{-1}\bigotimes_{j=1}^n P_j^{(2i)}U \bigotimes_{j=1}^n P_j^{(2i-1)}))^{-1}$ and the inverse of the local Clifford gate $\bigotimes_{i=1}^n C_i^{-1}$ are applied thereafter. Finally, one applies computational-basis measurements and records the outcome. One needs to sample $K_r$ random sequences, and for each sequence, one measures $K_s$ times. The outcome statistics are counted for each sequence, as in (b). After that, one randomly chooses $K_q$ observables $Z_w\in \{\mathbb{I}, Z\}^{\otimes n}$, where $w\in \{0,1\}^n$ and $\mathrm{pt}(Z_w) = w$, with probability $2^{-2n}3^\abs{w}$. One estimates the expectation values of these chosen observables, and the expectation values need to be averaged across $K_r$ random sequences. The above procedure is repeated for different $m$ from a circuit depth set, $\{m_1, m_2, \cdots, m_M\}$. For all $m$, the chosen observables have to be the same. (c) demonstrates the expectation values of different observables for different sequence lengths. The expectation value $O_w(m)$ is approximately proportional to $\lambda_w^{2m}$, which can be fit to $A\lambda^{2m}$ to determine quality parameter $\lambda_w$ like (d). (e) shows the last step. The final fidelity estimation is the average of the fitting values, and subsequently, one can further evaluate the gate correlation and employ gate optimization. Practical experimental settings choose constant values for $K_r$, $K_s$, and $K_q$ independent of the qubit count.}
\label{fig:cab}
\end{figure}

The core of CAB lies in assessing the quality parameter of the channel $\sqrt{U^{-1}\Lambda'_UU\Lambda_U}$ through the circuit depicted in Figure~\ref{fig:cab}(a), with $\Lambda_U$ and $\Lambda'_U$ being the Pauli-twirled noise channels of $U$ and $U^{-1}$, respectively. This approach yields the CAB fidelity of $U$, closely approximating $U$'s fidelity under physically reasonable conditions~\cite{Zhang2023cab}. Hereafter, we refer to the CAB fidelity simply as the gate fidelity unless stated otherwise. While estimating all quality parameters for fidelity evaluation demands exponential resources, the number of quality parameters necessary for accurate fidelity estimation within acceptable error margins and confidence levels is independent of the qubit count, thereby ensuring the scalability of the protocol. The whole procedure is shown in Figure~\ref{fig:cab} and elaborated in Methods. Note that the fidelity evaluated by this procedure is associated with the noise of $U$ and that of the local twirling gates adjacent to $U$ and $U^{-1}$, referred to as dressed fidelity. To isolate $U$'s fidelity, one can employ interleaved RB techniques~\cite{Magesan2012interleavedRB}, comparing the dressed fidelity against the local twirling gate fidelity. Particularly, the fidelity of $U$ is derived by~\cite{Magesan2012interleavedRB}
\begin{equation}\label{eq:interleave}
F = \frac{4^nF_{\mathrm{dress}}-1}{4^nF_{\mathrm{twirl}}-1}(1-\frac{1}{4^n})+\frac{1}{4^n},
\end{equation}
with $F_{\mathrm{dress}}$ and $F_{\mathrm{twirl}}$ the dressed and twirling gate fidelities, respectively.
The local twirling gate fidelity itself is determined through CAB, with the identity operation as the target gate.

\subsection{Fully connected gate and parallel CZ gate benchmarking}\label{ssc:fully}
\begin{figure}[!htbp]
\centering \resizebox{18cm}{!}{\includegraphics{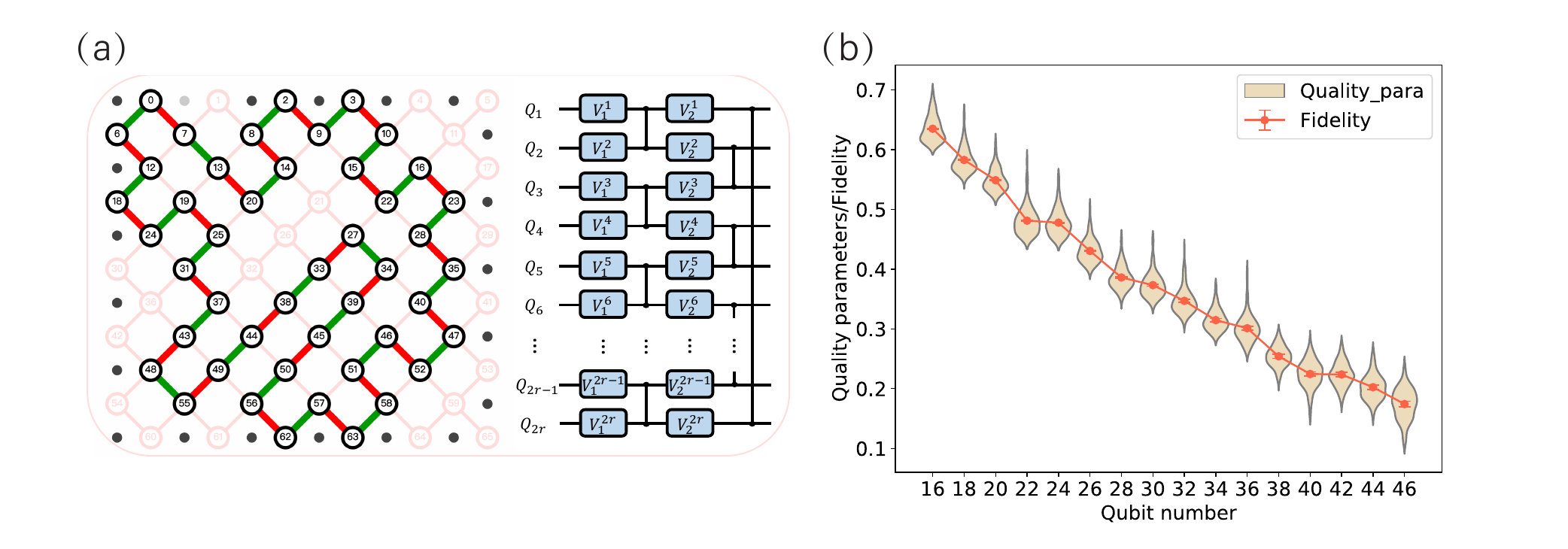}}
\caption{\textbf{The topological diagram and corresponding fidelity of a fully  connected gate.} (a) The left figure demonstrates the qubits and CZ gates used to realize the fully connected gate with 46 qubits. We select a ring on the two-dimensional quantum processor, corresponding to a one-dimensional quantum system. The ring contains two patterns of parallel CZ gates, shown in red and green, respectively. The right figure shows the circuit structure for the fully connected quantum gate on $n = 2r$ qubits, where $Q_i$ represents the $i$-th qubit. The gate comprises two layers of single-qubit gates, $\bigotimes_{i=1}^{n}V_1^{i}$ and $\bigotimes_{i=1}^{n}V_2^{i}$, and two layers of different parallel CZ gates. (b) The quality parameters and the fidelities of the dressed gates, including the error from the local twirling gates and the target gates. The distributions of the noise channel quality parameters are shown with violin plots. Their mean values equal the gate fidelities, shown with a point with an error bar. The length of the error bar equals the standard error of the fidelity estimation.
}
\label{fig:fig2}
\end{figure}

Our CAB experiments for the fully connected gate and the parallel CZ gate are conducted on a 54-qubit superconducting quantum computer. On a two-qubit system, the CZ gate is $\mathrm{CZ} = \ketbra{0}\otimes \mathbb{I} + \ketbra{1}\otimes Z$, and its parallel extension across multiple qubits is defined by $U = \bigotimes_{k=1}^{r} \mathrm{CZ}^{(i_k,j_k)}$, where $\mathrm{CZ}^{(i_k,j_k)}$ denotes the CZ gate acting on a specific qubit pair, $(i_k, j_k)$, making $U$ a $2r$-qubit gate. The fully connected gate comprises two layers of different parallel CZ gates intertwined with two layers of single-qubit gates, as shown in Figure~\ref{fig:fig2}(a). One of the parallel CZ gates connects qubits 1 and 2, 3 and 4, ... and the other connects qubits 2 and 3, 4 and 5...

\textit{Fully connected gate benchmarking.}---The single-qubit gates in the fully connected gate, like $V_1^i$ and $V_2^i$ in Figure~\ref{fig:fig2}(a), are free to vary. In quantum information tasks like variational quantum algorithms, they are normally chosen according to specific problems. In our experiments, we randomly sample all the single-qubit gates from the Clifford group so that the global gate is a Clifford gate and can be benchmarked with CAB. It is worth mentioning that cycle benchmarking requires repeatedly implementing the target gate $U$ proportional to the gate order, defined as the smallest positive integer $p$ to make $U^p = \mathbb{I}$. The fully connected gate generally has a rapidly increasing order with respect to the qubit number, and the order has already been thousands on average for 16 qubits, which we show in the Supplemental Material~\cite{supplementary}. The long-depth circuit will lead to extremely noisy experimental results and cannot provide effective benchmarking. Instead, by using the target gate and its inverse, this gate can be benchmarked within a short-depth circuit using CAB.

We realized the fully connected gate with qubit numbers from 16 to 46 and benchmarked its dressed fidelity. The dressed fidelity of each target gate is estimated with circuit depths of $\{0, 1\}$ through 50 circuit samples per circuit depth and performed 20,000 measurements per circuit. Theoretically, larger circuit depth differences can reduce fidelity estimation fluctuations, but practical constraints limit how deep the circuits can be. To reduce the impact of gate noise on the measurement results, we choose the circuit depths below 2.

Based on the measurement results, we estimated 100 quality parameters and averaged them to compute the gate fidelity. The results are shown in Figure~\ref{fig:fig2}(b), ranging from $63.49\%\pm 0.07\%$ to $17.42\%\pm 0.45\%$ for qubit numbers from 16 to 46, with the full data available in the Supplemental Material~\cite{supplementary}. The quality parameters are distributed near the mean value, meaning the noise is close to a depolarizing noise. This feature is extremely useful in realizing CAB experiments in a low-fidelity region. If the noise is far from the depolarizing noise, like the unitary noise, and the fidelity is low, the quality parameters may be below 0. Since a quality parameter $\lambda$ is obtained by fitting it to $A\lambda^{2m}$ as shown in Figure~\ref{fig:cab}, a negative quality parameter $\lambda$ is indistinguishable from $-\lambda$ and cannot be obtained correctly by the exponential fitting.

\begin{figure}[!htbp]
\centering \resizebox{17cm}{!}{\includegraphics{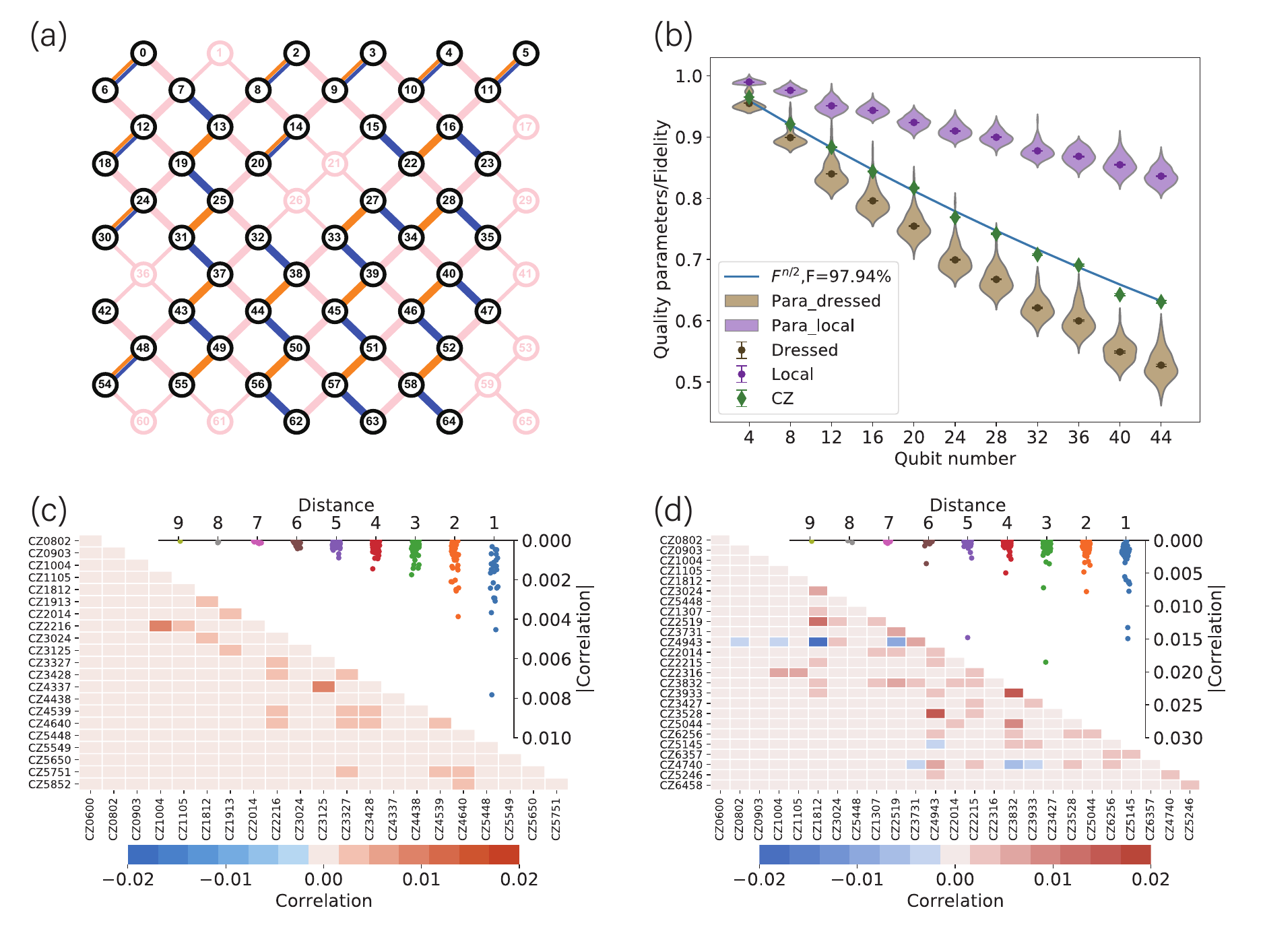}}
\caption{\textbf{The topological diagram and corresponding experimental results of  parallel CZ gate benchmarking.} (a) Two parallel CZ gate patterns, colored orange (22 CZ gates) and blue (26 CZ gates) on a 54-qubit quantum processor, with available qubits shown in black. (b) Benchmarking fidelities for the parallel CZ gate within the orange pattern. The benchmarking is done progressively from 2 to 22 CZ gates. Brown and purple violin plots illustrate the noise channel quality parameters for the dressed and local twirling gates, respectively. The point with an error bar in each violin plot represents the mean value equal to the gate fidelity, and the length of the error bar equals the standard error of the fidelity estimation.  Green diamond-shaped points denote individual fidelities of target gates, fitting well with the theoretical curve $F^{n/2}$, where $n$ is the qubit number and $F=97.94\%$. (c) A heatmap displays the pairwise correlation among CZ gates within the orange pattern, revealing weak positive correlations among several neighboring gates. We label the indexes of the CZ gates on the two axes. CZ$AB$ means the CZ gate between qubits $A$ and $B$. For instance, CZ$0802$ is the CZ gate between qubits $2$ and $8$. The scatter diagram shows that the absolute value of the correlation decreases with the distance of the CZ gates and is mainly large between nearby CZ gates. The CZ gate distance is measured as the minimal line count connecting qubit pairs in the processor layout from (a). (d) The heatmap and scatter plot for the blue pattern, with the same color scale and interpretation as in (c), indicate more significant correlations compared to the orange pattern and show that large correlations also exist for two remote CZ gates.}
\label{fig:fig3}
\end{figure}

\textit{Parallel CZ gate benchmarking.}---The benchmarking for the parallel CZ gates contains two patterns, depicted in Figure~\ref{fig:fig3}(a) using two distinct colors. We first evaluate the fidelities of the gates within the orange pattern, consisting of 22 pairs of CZ gates aligned in the same physical direction. This benchmarking was conducted progressively, starting with 2 pairs of CZ gates and incrementally including more gates up to the full set of 22 pairs. Subsequently, we evaluate the gate correlations within the orange and the blue patterns. The latter incorporates 26 pairs of CZ gates and engages almost the entire quantum processor.

The settings for benchmarking parallel CZ gates within the orange pattern are the same as that for the fully connected gate, except the circuit depths changed to $\{0, 2\}$. In this experiment, we benchmark both the dressed fidelity and the local twirling gate fidelity. The local twirling gate fidelity is obtained by changing the target gate $U$ with the identity operation. We then use the interleaved technique~\cite{Magesan2012interleavedRB} and Eq.~\eqref{eq:interleave} to isolate the pure parallel CZ gate fidelity.

In Figure~\ref{fig:fig3}(b), we show the noise channel quality parameter distribution for the dressed parallel CZ gates and the local gates with violin plots. The mean value equals the fidelity and is shown with a point. The green diamond-shaped points represent the pure fidelities of the parallel CZ gates within the orange pattern. The largest one, or the 22-pair parallel CZ gate, possesses a fidelity of $63.09\%\pm 0.23\%$. These fidelities have been fit using the function $F^{n/2}$, where $n$ is the qubit number. The fit aligns closely with our experimental data, suggesting a fidelity value of approximately $97.94\%$ for a single CZ gate. This indicates that, in the orange pattern, the fidelity of individual CZ gates remains nearly constant and is not affected by an increase in qubit number. This observation implies that the crosstalk among these parallel CZ gates is either limited to short-range interactions or is remarkably minimal. Such a characteristic is critical for the implementation of quantum error correction. The detailed fidelity and standard error data are available in the Supplemental Material~\cite{supplementary}.

\subsection{Correlation benchmarking}\label{ssc:correlation}
Beyond gate fidelity, our analysis extends to examining correlations within parallel CZ gates. We define this correlation as the deviation of the parallel CZ gate fidelity from the product of the fidelities of its comprising local CZ gates. The concept of correlation is elaborated in Methods. Any nonzero correlation value emerges as an indicator of interactions among the local gates.

The correlations among every two CZ gates within the orange and blue patterns are visualized through heatmaps in Figures~\ref{fig:fig3}(c) and (d), respectively. Additionally, we plot the correlation magnitudes as a function of the physical distance between CZ gates in the processor. For the orange pattern, we used the pure CZ gate fidelities from the parallel CZ gate benchmarking experiment to evaluate the correlation. The result of the blue pattern is obtained by another experiment.
The orange pattern exhibits a notable feature: correlations between CZ gates are consistently positive and pronounced only when the gates are close. This observation suggests a negligible presence of long-range interactions in the implementation. However, the 26-pair parallel CZ gate within the blue pattern reveals different performance, with substantially higher correlations even for distant CZ gates, indicating the existence of long-range interactions within this configuration. The difference between the two patterns can be explained by whether a parallel calibration of CZ gates before benchmarking exists. Before experiments, gates in the same direction--such as all the two-qubit gates in the orange pattern--were calibrated in parallel using the method detailed in the ``Parallel calibration of controlled-Z gate parameters with back probability" subsection in Methods. To maximize the qubit number of a pattern, the blue pattern is formed by combining gates in two directions. The results show that correlation benchmarking serves as an additional indicator of gate performance, alongside fidelity.

In Methods, we establish a noise model composed of depolarizing and $ZZ$-coupling noises to explain the correlation benchmarking results. The model first introduces a local depolarizing noise on each gate, followed by a unitary correlated noise. The Hamiltonian of the correlated noise is a summation of pairwise $ZZ$ on each pair of CZ gates, where the coefficients depend on the coupling strengths. Our analysis focuses on the two-gate and three-gate coupling cases. A natural result is that the correlation value between two gates positively depends on their own coupling strength. Weak correlation implies weak coupling. Interestingly, we find that when only two gates couple with each other, the correlation is always positive. The correlation value is normally less than 0.001 for uncorrelated gates and can be on the order of 0.01 for correlated gates, which is consistent with the experimental data. Nonetheless, if one of the two gates couples with a third gate, the correlation between the original two gates can become negative. Particularly, when one gate is strongly coupled with the third gate, and the other gate is weakly coupled to it, the negative correlation becomes significant.

Applying the theoretical analysis to the experimental data, we observe that most correlation values are positive, corresponding to weak coupling or two-gate coupling cases. The results of the orange pattern can be fully explained by two-gate coupling. The pair of CZ2216 and CZ1004 and the pair of CZ4337 and CZ3125 exhibit the strongest couplings. The CZ gates in the two pairs are both nearest neighbors. For the blue pattern, negative correlation values are observed, indicating that the two-gate coupling model is not sufficient to explain the data. Take the pair of CZ4943 and CZ1812 as an example, the negative correlation value implies that besides the coupling between CZ4943 and CZ1812, there exists a third gate strongly coupled with CZ4943 or CZ1812. From the correlation data, we infer that a neighbor of CZ4943, like CZ3731, couples strongly with CZ4943 but not with CZ1812. The coupling relationship among these three CZ gates can lead to a negative correlation between CZ4943 and CZ1812. Note that in real experiments, couplings can involve more than three gates, making the origins of negative correlations more complicated than the three-gate model described here. We expect our results to inspire more explorations into the correlation and coupling among quantum gates.

\subsection{Parallel CZ gate optimization}\label{ssc:optimization}
In addition to benchmarking, we conducted optimization on parallel CZ gates, employing two distinct approaches: optimizing with global fidelity and optimizing with individual local CZ gate fidelities. In both cases, the Nelder-Mead algorithm is utilized for optimization~\cite{nelder1965simplex, rol2017restless}.
Note that the noise of the target gate is much larger than that of the local twirling gates, and the fidelities of the local twirling gates are stable. The noise of the target gate dominates the dressed fidelity, making this quantity sufficient for optimization. To reduce benchmarking time, we optimize using the dressed fidelity, avoiding additional benchmarking of local twirling gates and the interleaved procedure. To minimize the influence of other unstable factors on gate fidelity, we measure both the fidelity of the parameters being iterated (iterative fidelity) and the fidelity of the initial parameters as a reference (reference fidelity) throughout the optimization process. The optimization target function is defined as the difference between these two fidelities. Before this optimization experiment, each local CZ gate was calibrated with the ``fast calibration" and ``parallel calibration" approaches shown in Methods.

\begin{figure}[!htbp]
\centering \resizebox{17cm}{!}{\includegraphics{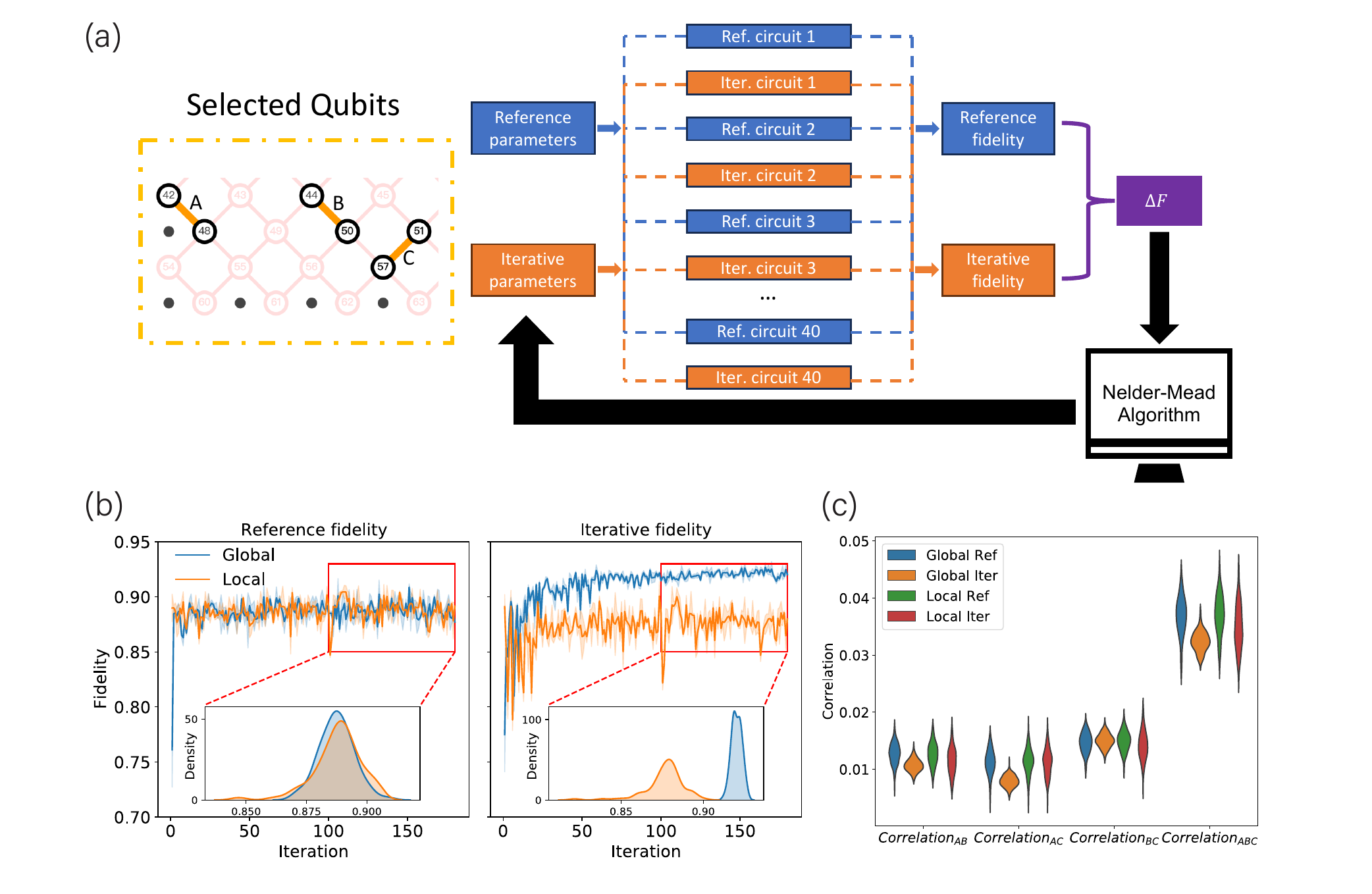}}
\caption{\textbf{The experimental optimization process and corresponding results of parallel CZ gates.} (a) The left figure shows the topology of 3-pair parallel CZ gates on the quantum processor. The three CZ gates are labeled with $A$, $B$, and $C$. During the optimization, we benchmark the fidelities of the target gate associated with two sets of parameters via CAB. The reference fidelity, fluctuating due to environmental factors, is based on reference parameters, and the iterative fidelity, mainly influenced by varying CZ gate parameters, corresponds to iterative parameters. The target function subtracts the reference fidelity from the iterative fidelity, thus mitigating the interference of environmental factors. The optimization algorithm is the Nelder-Mead algorithm.
(b) The data of global fidelity during the optimization procedure. The line is the fidelity estimation, and the shadow above and below represents the value of one standard error away from the fidelity. The blue and orange lines correspond to the optimization results utilizing global and local CZ gate fidelities, respectively. The left and right figures show the reference and iterative fidelities, respectively. The small figure shows the probability density distribution of the fidelities within iterations 100-180, obtained by kernel density estimation. During this range, two reference fidelities remain stable and close, and iterative fidelities converge. Comparing the iterative fidelities in this range allows for a fair assessment, with the right figure indicating more effective fidelity improvements when global fidelity is used for optimization. (c) The distribution of gate correlations derived from reference and iterative fidelities within iterations 100-180 in (b). For a 3-pair parallel CZ gate, correlations are observed among all three gates ($Correlation_{ABC}$) and between each pair ($Correlation_{AB}$, $Correlation_{AC}$, and $Correlation_{BC}$). For instance, $Correlation_{AB}$ refers to the correlation between CZ gates $A$ and $B$. The average values and standard deviations of correlations from reference fidelities are similar for both objective functions, indicating consistent environmental influences. However, for iterative fidelities, except $Correlation_{BC}$, employing global fidelity as the target function generally leads to improved outcomes.}
\label{fig:opt3CZ}
\end{figure}

In our experiments, the parallel CZ gate comprises $2n$ optimizable parameters with $n$ the qubit number. We chose $n$ as 4 and 6 so that the number of parameters is suitable for the Nelder-Mead algorithm to work. Optimizing gates with tens of qubits requires more scalable optimization algorithms.
Figure~\ref{fig:opt3CZ} presents the optimization results for a parallel CZ gate comprising 3 local CZ gates on 6 qubits. The topology of the three CZ gates and the optimization procedure are shown in Figure~\ref{fig:opt3CZ}(a). The three gates are relatively close, which are more likely to correlate with each other. Meanwhile, the readout channels of these three gates are relatively stable compared to other gates, ensuring minimal influence of environmental noise on the optimization procedure.

The progression of fidelities during the optimization is depicted in Figure~\ref{fig:opt3CZ}(b), and the inter-gate correlations, calculated based on data from iterations 100-180, are illustrated in Figure~\ref{fig:opt3CZ}(c). This specific iteration range is chosen as it is the phase of the iterative parameter convergence and stable reference fidelities, indicating a reduced impact from other fluctuating factors. Further experimental details and results of a 4-qubit setup are available in the Supplemental Material~\cite{supplementary}.

Below, we compare the optimization results employing global fidelity and the local gate fidelities with data from iterations 100-180.
The fidelity is improved to 92.04\% and 87.65\%, and the correlation is reduced to 3.22\% and 3.53\% when using global fidelity and the local gate fidelities for optimization, respectively.
It is clear that using global fidelity for optimization more effectively enhances fidelity and reduces correlation. Using local gate fidelities tends to yield inferior optimization outcomes, attributable to the lack of correlation information within the target function. In Methods, we use the $ZZ$-coupling model to explain the difference between global and local gate fidelities. When three gates are mutually coupled, optimizing one local fidelity may cause a decrease in the other two. This antagonistic relationship between local fidelities can trap the optimization in a local region. In contrast, global fidelity incorporates all correlation information, allowing the optimization to proceed monotonically.
This finding underscores the significance of correlation in optimizing parallel gates and demonstrates the crucial advantage and essence of benchmarking large-scale quantum gates.

\section{Discussion}\label{sc:discussion}
In conclusion, we utilize CAB to conduct large-scale experiments of benchmarking the fully connected and parallel CZ gates. The benchmarking of gate correlation provides quantity to evaluate the quantum gate performance beyond gate fidelity, allowing the detection of long-range interaction, which may further be useful in studying many-body physics. Combined with the $ZZ$-coupling noise model, it is feasible to detect the coupling pattern of the parallel gates with correlation benchmarking. The established noise model also provides insight into further study of near-term quantum devices and quantum error correction.

The results highlight the crucial role of correlation in optimizing parallel quantum gates. Practically, one can decompose the circuit into multiple layers and optimize each layer with improved gate fidelity and reduced inter-gate correlation. This approach is more effective than optimizing each local gate individually, as evidenced by our optimization results. Meanwhile, one can first identify strongly coupled gates through correlation benchmarking and divide them into distinct groups. Gates within each group are strongly correlated, while groups themselves are weakly correlated. By optimizing each group individually, the overall performance of the entire layer can be enhanced. This approach simplifies the optimization process by focusing on the dominant gate correlations.

Our experimental methodology for benchmarking and optimizing large-scale gates applies to enhancing all Clifford circuits up to a local gauge transformation~\cite{Zhang2023cab}, including the essential case of quantum error-correcting code circuits~\cite{gottesman1997stabilizer}. Typical quantum encoding schemes like surface codes~\cite{Zhao2022Correcting,Acharya2023Suppressing} or, more generally, quantum low-density-parity-check codes~\cite{Bravyi2024Highthreshold}, involve tens or even hundreds of qubits. Optimizing such large batches involves managing many gate parameters, necessitating the development of scalable optimization algorithms rather than solely scalable benchmarking. In the future, advanced large-scale optimization algorithms, particularly gradient-free ones, will help to explore large-gate optimization, ultimately contributing to realizing universal fault-tolerant quantum computers.

\begin{acknowledgements}
We thank Pengyu Liu, Yuxuan Yan, and Zhiyuan Chen for the helpful discussions. This work was supported by the Innovation Program for Quantum Science and Technology (Grant No.~2021ZD0300200), Anhui Initiative in Quantum Information Technologies, Special funds from Jinan Science and Technology Bureau and Jinan high tech Zone Management Committee, Shanghai Municipal Science and Technology Major Project (Grant No.~2019SHZDZX01), the New Cornerstone Science Foundation through the XPLORER PRIZE, the Key-Area Research and Development Program of Guangdong Province (Grant No.~2020B0303060001), Shanghai Science and Technology Development Funds (Grant No.~23YF1452500), the National Natural Science Foundation of China (Grant No.~12174216), the Innovation Program for Quantum Science and Technology (Grant No.~2021ZD0300804).
\end{acknowledgements}

\section{Methods}
\subsection{Procedure of character-average benchmarking}\label{methodsc:cab}
Below in Box~\ref{box:cab}, we introduce the procedure of the CAB protocol when benchmarking an $n$-qubit Clifford gate corresponding to Figure~\ref{fig:cab}. For target gates as non-Clifford gates and more protocol details, one can refer to Ref.~\cite{Zhang2023cab}.
\begin{mybox}{box:cab}{Procedure of character-average benchmarking}
\begin{enumerate}
\item Choose a list of circuit depths $\{m_1, m_2, \cdots, m_M\}$, where $M$ is the number of circuit depths.
\item Choose integers $K_r$ and $K_s$ as the number of random sequences and single-shot measurements, respectively.
\item For any $1\leq j\leq M$, choose $K_r$ random sequences $\mathcal{S}_j^k = C^{-1}U_{inv}\prod_{i=1}^{m_j}(U^{-1}P^{(2i)}UP^{(2i-1)})C$. Here, $C$ is a local Clifford gate uniformly and randomly sampled from the $n$-qubit local Clifford group $\textsf{C}_1^{\otimes n}$, and $\forall 1\leq i\leq 2m_j$, $P^{(i)}$ is uniformly and randomly sampled from the $n$-qubit Pauli group $\textsf{P}_n$. The inverse gate $U_{inv} = (\prod_{i=1}^{m_j}(U^{-1}P^{(2i)}UP^{(2i-1)}))^{-1}$ is a Pauli gate.
\item Prepare state $\ket{0}^{\otimes n}$, implement each random sequence $K_s$ times, and collect all $Z$-basis measurement results.
\item\label{item:cabmodify} Independently sample $K_q$ $Z$-basis observables, $\{O_i, 1\leq i\leq K_q\}$, from $\{\mathbb{I}, Z\}^{\otimes n}$. The sampling distribution is given by $2^{-2n}3^{|O_i|}$, where $|O_i|$ is the weight of $O_i$, or the number of $Z$ in $O_i$. For each $O_i$, compute $\tr{\widetilde{O}_i\mathcal{S}_j^k(\rho)}$ with the measurement results from the previous step where $\rho$ is the noisy version of $\ketbra{0}^{\otimes n}$, and $\widetilde{O}_i$ is the noisy version of $O_i$. Then, average $\tr{\widetilde{O}_i\mathcal{S}_j^k(\rho)}$ over different random sequences for each circuit depth $m_j$ and obtain $f_i(m_j) = \frac{1}{K_r}\sum_{k=1}^{K_r} \tr{\widetilde{O}_i\mathcal{S}_j^k(\rho)}$, which is named survival probability.
\item For each $O_i$, fit $\{f_i(m_j), m_j\}$ to the function $f_i(m) = A\lambda_i^{2m}$ and obtain $\lambda_i$, which is a quality parameter of the noise channel. Then, the process fidelity of the target gate $U$ is given by the average of $\{\lambda_i, 1\leq i\leq K_q\}$,
\begin{equation}\label{eq:processfidelityest}
F = \frac{1}{K_q} \sum_{i=1}^{K_q} \lambda_i.
\end{equation}
\end{enumerate}
\end{mybox}

Note that the procedure in Box~\ref{box:cab} differs from the original one in Ref.~\cite{Zhang2023cab}. The main modification lies in step~\ref{item:cabmodify}. In Ref.~\cite{Zhang2023cab}, one does not sample observables but take all observables from $\{\mathbb{I}, Z\}^{\otimes n}$, which requires $K_q = 2^n$. Here, we only need to set $K_q = O(-\epsilon^{-2}\log \delta)$ to ensure that Eq.~\eqref{eq:processfidelityest} only differs from the fidelity estimated by traversing observables a small quantity, $\epsilon$, with a high confidence level, $1-\delta$. This can be seen from Hoeffding's inequality, which is shown below. Note that $\lambda_i$ is limited in the region $[-1, 1]$.
\begin{equation}
\Pr( |F-\mathbb{E}_{\lambda_i} [F] | > \epsilon) \leq 2\exp(-\frac{K_q\epsilon^2}{2}).
\end{equation}
Setting $2\exp(-\frac{K_q\epsilon^2}{2}) = \delta$, we get $K_q = 2\epsilon^{-2}(\log 2\delta^{-1})$. Note that $K_q$ is irrelevant to the qubit number $n$. Thus, the complexity of the classical postprocessing is independent of $n$. The number of sampled sequences for fidelity estimation is also independent of $n$ as proved in Ref.~\cite{Zhang2023cab}. Thus, the whole benchmarking protocol is scalable.

\subsection{Correlation}\label{methodsc:correlation}
Here, we introduce the formal definition of the correlation of a parallel gate. We consider a parallel gate, $U = \bigotimes_{i=1}^g U_i$, where $U_i$ is more local or acts on fewer qubits than $U$. Normally, $U_i$ is a one-local or two-local gate in experiments. That is, $U_i$ only acts on one qubit or two qubits. Via CAB, one can simultaneously get the fidelity of $U$ and the fidelities of $U_i$, denoted as $F(U)$ and $F(U_i)$, respectively. If there is no correlation among $U_i$, the noise channel of $U$ can be expressed as $\Lambda=\bigotimes_{i=1}^g \Lambda_i$ where $\Lambda_i$ is the individual noise of $U_i$. Then, the global fidelity, $F(U)$, would be equal to the product of local gate fidelities, $\prod_{i=1}^g F(U_i)$. In reality, the interaction among different gates would make the two values different. We define the following quantity to characterize the total correlation among $\{U_i, 1\leq i\leq g\}$,
\begin{equation}\label{eq:correlation}
C_g(U, \{U_i, 1\leq i\leq g\}) = \frac{F(U)-\prod_{i=1}^g F(U_i)}{\sqrt{F(U)\prod_{i=1}^g F(U_i)}}.
\end{equation}
The denominator is a normalization factor. When the correlation is positive, the global fidelity is larger than the product fidelity, indicating that the correlation helps to increase the global fidelity. Since Eq.~\eqref{eq:correlation} is defined among $g$ gates, we call it $g$-correlation. Except for $g$-correlation, one can also obtain $j$-correlation among each $j$ gates in $\{U_i, 1\leq i\leq g\}$ where $2\leq j\leq g-1$ for parallel gate $U = \bigotimes_{i=1}^g U_i$. Note that since $F(U)$ and $F(U_i)$ can be obtained simultaneously from the same experimental data, gate correlation can also be evaluated concurrently without any additional experimental effort.

\subsection{Experimental platform}
In this work, we utilized a processor with the same design as the $Zuchongzhi 2.0$ processor~\cite{YulinWu2021Superconducting}, and selected up to 54 qubits for our experiments. The basic performance of the processor is shown in Table~\ref{tab:SSP}, where the single-qubit gate error and the two-qubit CZ gate error with a median of $0.24\%$ and $3.21\%$ by cross-entropy benchmarking (XEB)~\cite{Barends0Diabatic, 2017Characterizing, Arute2019}. Our scheme to realize a two-qubit CZ gate is implementing an all-microwave coupler with a fixed gate time of 110 ns~\cite{2021Realization}. The approach involves applying a microwave signal with an envelope $A(t)$ and a driving frequency of $\omega_t$ to the tunable coupler, with an extra flux given by $\Phi(t) = A(t)\cos(\omega_t t + \phi_0)$.  When the driving frequency $\omega_t$ matches the energy difference between the $\ket{11}$ and $\ket{02}$ states, i.e., $\omega_t = \omega_{11}-\omega_{02}$, resonance takes place between these two states. Here, $\ket{11}$ is a computational-basis state with each qubit at state $\ket{1}$, and $\ket{02}$ is a state outside the computational subspace. However, due to the nonlinear relationship between the extra flux and the coupling strength, although $\Phi(t)$ is a good single-frequency signal when transformed into the coupling strength, the signal contains significant frequency components not only at $\omega_t$, but also at $2\omega_t$ and $4\omega_t$. Therefore, when considering the frequency layout of qubits, it is necessary to avoid $\omega_t$, $2\omega_t$, and $4\omega_t$ equal to either $\Delta_{01,10} = \abs{\omega_{01}-\omega_{10}}$ or $\Delta_{11,20}=\omega_{11}-\omega_{20}$.

\begin{table}[!htbp]
\centering
\caption{Summary of system parameters. SD denotes the standard deviation.}
\begin{tabular}{lccc}\hline
Parameters & Mean & Median & SD  \\\hline
Qubit maximum frequency (GHz) & 5.504 & 5.504 & 0.093 \\
Qubit idle frequency (GHz) & 5.411 & 5.403 & 0.104 \\
Qubit anharmonicity (MHz) & -243 & -242 & 4 \\
$T_1$ at idle frequency ($\mu$s) & 22.14 & 22.11 & 7.60 \\
$T_2^*$ at idle frequency ($\mu$s) & 4.94 & 4.55 & 3.24 \\
Readout $e_{|0\rangle}$ ($\%$) & 1.03 & 0.90 & 0.72 \\
Readout $e_{|1\rangle}$ ($\%$) & 3.82 & 3.43 & 1.64 \\
1Q XEB $e_1$ ($\%$) & 0.31 & 0.24 & 0.18\\
2Q-CZ XEB $e_2$ ($\%$) & 3.35 & 3.21 & 0.98 \\\hline
\end{tabular}
\label{tab:SSP}
\end{table}

Before our experiments, we adjusted the qubit frequency and calibrated the parameters for each local gate to make the processor perform well. In the following subsections, we will elaborate on this procedure in detail.

\subsection{Frequency conflict and frequency adjustment}
The distribution of qubit frequencies is typically constrained within a range of 0-400 MHz due to magnetic flux noise and the bandwidth of the digital-to-analog converter. When arranging the qubit frequencies, the following factors need to be considered and balanced:
(1) Energy relaxation time $T_1$ and dephasing time $T_\phi$.
(2) Spacing of frequencies between neighboring qubits and next-to-nearest neighboring qubits.
(3) Two-qubit gate frequencies, as well as their second and fourth harmonic frequencies, and the frequency conflicts with $\Delta_{01,10} = |\omega_{01} - \omega_{10}|$ and $\Delta_{11,20} = \omega_{11} - \omega_{20}$.
(4) Maximum frequencies for each qubit, which represent the available frequency range for each qubit. 
By defining the above factors as error functions and frequency domains, we can obtain a set of theoretically optimal frequencies. After adjusting the frequencies of all qubits to the optimized arrangement, a majority of single-qubit gates and two-qubit gates can achieve high fidelity through standard calibration. However, local fine-tuning is still required for poorly performing gates. Additionally, the performance of a quantum processor may deteriorate during certain time intervals due to long-term periodic frequency variations in two-level systems. This also requires fine-tuning of the corresponding qubits. Qubit frequency tuning is relatively frequent and tedious, as adjusting the frequency of one qubit requires re-calibrating two-qubit gates associated with it. Therefore, it is crucial to calibrate single-qubit gates and two-qubit gates efficiently in this process. On the other side, the success of subsequent benchmarking relies on an initial good adjustment of qubit frequencies, as a higher gate fidelity improves the benchmarking accuracy and stability. The configuration of qubit frequencies is the key to the success of our experiments.

\subsection{Fast calibration of controlled-Z gates}\label{methodsc:fast}
\begin{figure}[!htbp]
\centering \resizebox{17cm}{!}{\includegraphics{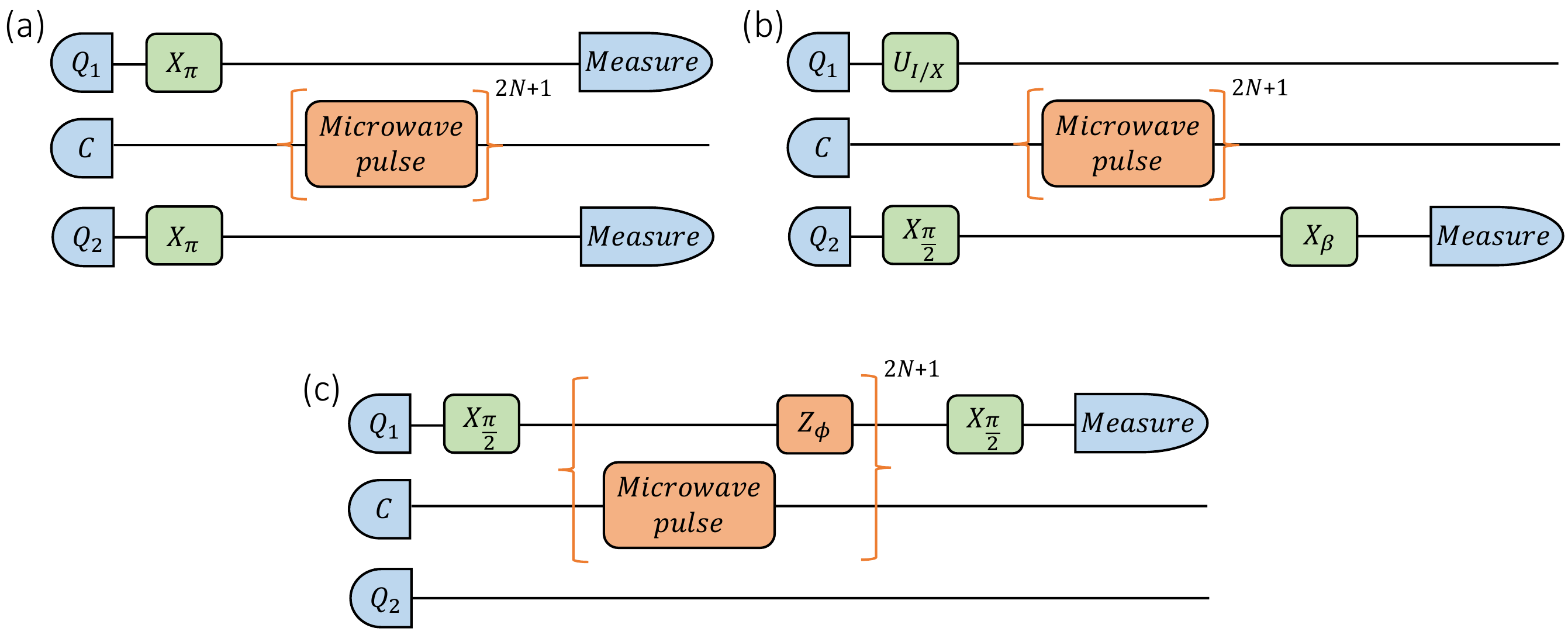}}
\caption{\textbf{Circuits for fast calibration of CZ gates.} $Q_1$ and $Q_2$ represent the first qubit and the second qubit, respectively. $C$ represents the coupler. $X_{\beta} = e^{-\frac{i\beta X}{2}}$ where $X$ is the Pauli $X$ operator. $N$ normally takes 1, 3, 5, 7.}
\label{fig:calibrationCircuit}
\end{figure}

After fine-tuning the qubit frequency, we need to recalibrate CZ gates related to it. The main parameters for calibrating the CZ gates are microwave frequency, microwave amplitude, and dynamic phase of the two relevant qubits. First, we roughly determine the microwave frequency and amplitude through the circuit in Figure~\ref{fig:calibrationCircuit}(a) with $N$ typically set to 0.
Within this circuit, the two qubits $Q_1$ and $Q_2$ are initially set at $\ket{0}$. We first flip these two qubits by applying pulse $X_{\pi}$. Then, we apply the microwave pulse once, and after that, we measure the probability of two qubits returning to the $\ket{11}$ state. In the process of applying microwave pulses, $\ket{11}$ and $\ket{02}$ states will be exchanged, and we try to find the microwave pulse parameters to maximize the probability back to $\ket{11}$ for the ending state.
Then, we fine-tune the microwave amplitude by implementing circuit (a) again. To amplify the errors caused by the parameters, we superimpose $2N+1$ CZ gates, with $N > 0$ this time. As the conditional phase is relatively sensitive to the frequency, we fine-tune the microwave frequency through the circuit in Figure~\ref{fig:calibrationCircuit}(b). When the $\beta$ of $X_\beta$ changes, the probability of $Q_1$, or the first qubit, changes as follows: 
\begin{align}
&P(\beta)=\frac{1+\cos(\beta+\phi_I)}{2}\ \mathrm{for}\  U = I; \\
&P(\beta)=\frac{1+\cos(\beta+\phi_x)}{2}\ \mathrm{for}\ U = X. 
\end{align}
By fitting $P(\beta)$ and obtaining $\phi_I$ and $\phi_x$, we can get the conditional phase $\phi=\phi_x-\phi_I$. The optimal microwave frequency is the frequency at which $\phi=\pi$ is satisfied. We then repeat the circuit (a) again to calibrate the microwave amplitude further with a large $N$. The optimal microwave amplitude is the amplitude that makes the probability of the $\ket{11}$ state closest to 1. The dynamic phase of the two relevant qubits can be calibrated through the circuit in Figure~\ref{fig:calibrationCircuit}(c). We first change the dynamic phase $Z_\phi$ of $Q_1$ and find the point where the probability of the $\ket{1}$ state is closest to 1 to complete the dynamic phase compensation for $Q_1$. This process is repeated for $Q_2$ subsequently with $X_{\frac{\pi}{2}}$ and $Z_\phi$ applied at $Q_2$ in circuit (c).

\subsection{Parallel calibration of controlled-Z gate parameters with back probability}\label{methodsc:opt}
When we calibrate CZ gate parameters by means of amplifying errors through the superposition of multiple layers of CZ gate circuits, we can attain a high level of fidelity for most CZ gates. To further refine the gate parameters, we resort to the Nelder-Mead algorithm to continue to search for CZ gate parameters.
Particularly, for each local CZ gate, we input $\ket{00}$, implement a random sequence of two-qubit Clifford gates, and record the probability of the final state back to $\ket{00}$. Each two-qubit Clifford gate is decomposed into CZ and single-qubit gates in implementation. Additionally, we alternate running the same set of random circuits with the reference and iterative parameters. The difference between their outcomes is used as the target function of the Nelder-Mead algorithm to mitigate environmental influence. To quickly calibrate a set of CZ gates where no two gates share a common qubit, we execute the above procedure in parallel for each local CZ gate. Each gate is calibrated independently using its own back probability. This parallel calibration method typically yields better gate parameters than the parameter scanning approach described in the previous subsection.

\subsection{Fidelity and correlation analysis of depolarizing and correlated noise}
In this part, we establish a simple but physical noise model to explain the experimental results. Note that CAB faithfully evaluates the process fidelity of quantum gates~\cite{Zhang2023cab}. We consider how the process fidelity behaves under a combination of local depolarizing noise and gate interaction. Particularly, for a unitary gate $U = \bigotimes_{i=1}^g U_i$, we consider the following noise model.
\begin{equation}
\Lambda = \Lambda_V\circ \bigotimes_i \Lambda_{p_i},
\end{equation}
where $\Lambda_{p_i}$ is a depolarizing noise on the $i$-th gate with parameter $p_i$ such that
\begin{equation}
\Lambda_{p_i}(\rho) = p_i\rho + (1-p_i)\frac{\mathbb{I}_i}{d_i}.
\end{equation}
Here, $d_i$ is the dimension of gate $U_i$, and $\mathbb{I}_i$ is the identity operator on this subsystem. The noise $\Lambda_V$ is a correlated noise among all local gates $U_i$, modeled as a unitary evolution:
\begin{equation}
\Lambda_V(\rho) = V\rho V^{\dagger}.
\end{equation}

The form of $\Lambda$ incorporates both decoherence on individual gates and interactions between gates. The decoherence on each gate is set as a depolarizing type, which is a standard setting in studies of near-term quantum devices~\cite{Daniel2021Limitations,Aharonov2023Noisy,yan2023limitations,morvan2024phase}. Different from previous works, we introduce an additional interaction term $V$, which makes the noise model more physical. The form of $V$ depends on the implemented gate $U$, which will be specified later. When considering the action of $V$ on a subsystem $S$, the remainder of the system is treated as a maximally mixed state or a thermalized state at infinite temperature. That is, for state $\rho_S$ on subsystem $S$,
\begin{equation}
\Lambda_V|_S (\rho_S) = \tr_{\Bar{S}}V(\rho_S\otimes \frac{\mathbb{I}_{\Bar{S}}}{d_{\Bar{S}}}) V^{\dagger}.
\end{equation}
Here, $\Lambda_V|_S$ is the restriction of $\Lambda_V$ to $S$, $\Bar{S}$ is the complementary subsystem of $S$, $d_{\Bar{S}}$ is the dimension of $\Bar{S}$, and $\mathbb{I}_{\Bar{S}}$ is the identity operator on $\Bar{S}$.

Since the gate $U = \bigotimes_{i=1}^g U_i$ comprises $g$ gates $\{U_1, U_2, \cdots, U_g\}$, we use $[g] = \{1, 2, \cdots, g\}$ to denote the whole system. Given a subset $S\subseteq [g]$, we can represent a part of the gate $U$, $\bigotimes_{i\in S}U_i$, whose fidelity is given by $F(\Lambda_V|_S \circ \bigotimes_{i\in S}\Lambda_{p_i})$ and denoted as $F_S$. Note that the process fidelity has an expression $F(\Lambda) = \tr(\ketbra{\Phi^+}\Lambda(\ketbra{\Phi^+}))$ where $\ket{\Phi^+}$ is a maximally entangled state on two copies of the system. Through direct calculation, we have that
\begin{equation}\label{eq:fidelity}
F_S = \sum_{L\subseteq S} p^L(1-p)^{S\backslash L} \frac{d_L}{dd_S^2} \Vert \tr_L V \Vert_2^2,
\end{equation}
where $L$ is a subset of $S$, $S\backslash L$ is the complementary set of $L$ in $S$, $d_L$ and $d_S$ are the dimensions of subsystems $L$ and $S$, respectively, and
\begin{equation}
p^L = \prod_{i\in L}p_i, (1-p)^{S\backslash L} = \prod_{i\in S\backslash L}(1-p_i), \Vert \tr_L V \Vert_2^2 = \tr_{\Bar{L}}( \tr_L V \tr_L V^{\dagger} ).
\end{equation}
Note that $\Bar{L}$ is the complementary set of $L$ in $[g]$. From Eq.~\eqref{eq:fidelity}, we can obtain the global fidelity and local gate fidelities, and hence evaluate the correlation and investigate how fidelities depend on noise parameters.

In our experiments, we mainly consider the parallel CZ gate $U = \bigotimes_{k=1}^{r} \mathrm{CZ}^{(i_k,j_k)}$. Based on experimental observations, the correlated noise mainly arises from the $ZZ$ coupling among qubits. Particularly, we consider a simplified noise model where only one qubit from each CZ gate couples with each other. Without loss of generality, we set this qubit as $i_k$. Meanwhile, the Hamiltonian of the correlated noise only contains two local terms while the strength between $i_k$ and $i_l$ is set as $\gamma_{kl}$. Thus, the correlated noise $V$ is
\begin{equation}\label{eq:correlatednoise}
V = e^{-i \sum_{1\leq k<l\leq r} \gamma_{kl} Z_{i_k}Z_{i_l}}.
\end{equation}
The strength parameter is determined by $\gamma_{kl} = g_{kl}t$, where $t$ is the evolution time, and $g_{kl}$ is the coupling strength between two qubits. In our experiments, $t$ corresponds to the two-qubit gate time, which is 110ns. For two physically isolated qubits, their coupling strength is typically less than 0.3 MHz, resulting in $\gamma_{kl}$ less than 0.033 for two uncorrelated gates. For two coupled qubits, $\gamma_{kl}$ can be on the order of 0.1.

Substituting Eq.~\eqref{eq:correlatednoise} into Eq.~\eqref{eq:fidelity} gives the global fidelity and local CZ gate fidelities. We provide the results when $r = 2$ and $r = 3$, which relates to our correlation benchmarking and gate optimization results. More general cases can be straightforwardly obtained using Eq.~\eqref{eq:fidelity}.

When $r = 2$, the fidelities of the two local CZ gates are
\begin{align}
F_1 &= p_1\cos^2 \gamma_{12} + \frac{1-p_1}{4};\\
F_2 &= p_2\cos^2 \gamma_{12} + \frac{1-p_2}{4}.
\end{align}
The global fidelity of the two CZ gates is
\begin{equation}
F_{[2]} = (p_1p_2+\frac{p_1(1-p_2)}{4}+\frac{p_2(1-p_1)}{4})\cos^2 \gamma_{12} + \frac{(1-p_1)(1-p_2)}{16}.
\end{equation}
The above gives the correlation when only two CZ gates correlate with each other via the $ZZ$ coupling:
\begin{equation}\label{eq:correlationtwo}
\begin{split}
\frac{F_{[2]}-F_1F_2}{\sqrt{F_{[2]}F_1F_2}} = &\frac{p_1p_2 \cos^2\gamma_{12}\sin^2\gamma_{12}}{\sqrt{F_{[2]}F_1F_2}}\\
\overset{p_1,p_2\rightarrow 1}{\approx} &\sin \gamma_{12}\tan \gamma_{12}.
\end{split}
\end{equation}
In the limit that $p_1$ and $p_2$ are close to $1$, $\sqrt{F_{[2]}F_1F_2}$ is approximately $p_1p_2 \cos^3 \gamma_{12}$. Then, the correlation is approximately $\sin \gamma_{12}\tan \gamma_{12}$. Given the value of $\gamma_{12}$ as 0.033 and 0.1, the correlation values take 0.001 and 0.01, respectively. This is consistent with our experimental results.

When $r = 3$, we give the fidelities of the three local CZ gates, the fidelities of each pair of CZ gates, and the global fidelity.
\begin{align}
F_1 &= p_1(\cos^2 \gamma_{12}\cos^2 \gamma_{13} + \sin^2 \gamma_{12}\sin^2 \gamma_{13}) + \frac{1-p_1}{16};\\
F_2 &= p_2(\cos^2 \gamma_{12}\cos^2 \gamma_{23} + \sin^2 \gamma_{12}\sin^2 \gamma_{23}) + \frac{1-p_2}{16};\\
F_3 &= p_3(\cos^2 \gamma_{13}\cos^2 \gamma_{23} + \sin^2 \gamma_{13}\sin^2 \gamma_{23}) + \frac{1-p_3}{16};
\end{align}
\begin{equation}
\begin{split}
F_{12} =& p_1p_2(\cos^2\Vec{\lambda}+\sin^2\Vec{\lambda})+\frac{p_1(1-p_2)}{16}(\cos^2 \gamma_{12}\cos^2 \gamma_{13} + \sin^2 \gamma_{12}\sin^2 \gamma_{13})+\\
&\frac{(1-p_1)p_2}{16}(\cos^2 \gamma_{12}\cos^2 \gamma_{23} + \sin^2 \gamma_{12}\sin^2 \gamma_{23}) + \frac{(1-p_1)(1-p_2)}{256};\\
\end{split}
\end{equation}
\begin{equation}
\begin{split}
F_{13} =& p_1p_3(\cos^2\Vec{\lambda}+\sin^2\Vec{\lambda})+\frac{p_1(1-p_3)}{16}(\cos^2 \gamma_{12}\cos^2 \gamma_{13} + \sin^2 \gamma_{12}\sin^2 \gamma_{13})+\\
&\frac{(1-p_1)p_3}{16}(\cos^2 \gamma_{13}\cos^2 \gamma_{23} + \sin^2 \gamma_{13}\sin^2 \gamma_{23}) + \frac{(1-p_1)(1-p_3)}{256};\\
\end{split}
\end{equation}
\begin{equation}
\begin{split}
F_{23} =& p_2p_3(\cos^2\Vec{\lambda}+\sin^2\Vec{\lambda})+\frac{p_2(1-p_3)}{16}(\cos^2 \gamma_{12}\cos^2 \gamma_{23} + \sin^2 \gamma_{12}\sin^2 \gamma_{23})+\\
&\frac{(1-p_2)p_3}{16}(\cos^2 \gamma_{13}\cos^2 \gamma_{23} + \sin^2 \gamma_{13}\sin^2 \gamma_{23}) + \frac{(1-p_2)(1-p_3)}{256};\\
\end{split}
\end{equation}
\begin{equation}
\begin{split}
F_{[3]} =& (p_1p_2p_3+\frac{p_1p_2(1-p_3)}{16}+\frac{p_1p_3(1-p_2)}{16}+\frac{p_2p_3(1-p_1)}{16})(\cos^2\Vec{\lambda}+\sin^2\Vec{\lambda})+\\
&\frac{p_1(1-p_2)(1-p_3)}{256}(\cos^2 \gamma_{12}\cos^2 \gamma_{13} + \sin^2 \gamma_{12}\sin^2 \gamma_{13})+\\
&\frac{p_2(1-p_1)(1-p_3)}{256}(\cos^2 \gamma_{12}\cos^2 \gamma_{23} + \sin^2 \gamma_{12}\sin^2 \gamma_{23})+\\
&\frac{p_3(1-p_1)(1-p_2)}{256}(\cos^2 \gamma_{13}\cos^2 \gamma_{23} + \sin^2 \gamma_{13}\sin^2 \gamma_{23})+\\
&\frac{(1-p_1)(1-p_2)(1-p_3)}{4096}.
\end{split}
\end{equation}
Here, $\cos\Vec{\lambda} = \cos\gamma_{12}\cos\gamma_{13}\cos\gamma_{23}$ and $\sin\Vec{\lambda} = \sin\gamma_{12}\sin\gamma_{13}\sin\gamma_{23}$.
In this case, the correlation between the first and second local CZ gates is given by
\begin{equation}\label{eq:correlationthree}
\begin{split}
\frac{F_{12}-F_1F_2}{\sqrt{F_{12}F_1F_2}} = &\frac{p_1p_2 ((\cos^2\Vec{\lambda}+\sin^2\Vec{\lambda})-(\cos^2 \gamma_{12}\cos^2 \gamma_{13} + \sin^2 \gamma_{12}\sin^2 \gamma_{13})(\cos^2 \gamma_{12}\cos^2 \gamma_{23} + \sin^2 \gamma_{12}\sin^2 \gamma_{23}))}{\sqrt{F_{12}F_1F_2}}\\
= &\frac{p_1p_2 \cos^2\gamma_{12}\sin^2\gamma_{12}(\cos^2\gamma_{13}-\sin^2\gamma_{13})(\cos^2\gamma_{23}-\sin^2\gamma_{23})}{\sqrt{F_{12}F_1F_2}}\\
\overset{p_1,p_2\rightarrow 1}{\approx} &\frac{ \cos^2\gamma_{12}\sin^2\gamma_{12}(\cos^2\gamma_{13}-\sin^2\gamma_{13})(\cos^2\gamma_{23}-\sin^2\gamma_{23})}{\sqrt{(\cos^2\Vec{\lambda}+\sin^2\Vec{\lambda})(\cos^2 \gamma_{12}\cos^2 \gamma_{13} + \sin^2 \gamma_{12}\sin^2 \gamma_{13})(\cos^2 \gamma_{12}\cos^2 \gamma_{23} + \sin^2 \gamma_{12}\sin^2 \gamma_{23})}}.
\end{split}
\end{equation}
Note that when $p_1$ and $p_2$ are close to $1$, the first term in the formula of fidelities dominates, and we get the above approximation. It is interesting that the sign of the correlation depends on $(\cos^2\gamma_{13}-\sin^2\gamma_{13})(\cos^2\gamma_{23}-\sin^2\gamma_{23})$. A negative correlation value implies that one of $\gamma_{13}$ and $\gamma_{23}$ is larger than $\pi/4$. 

\begin{figure}[!htbp]
\centering \resizebox{16cm}{!}{\includegraphics{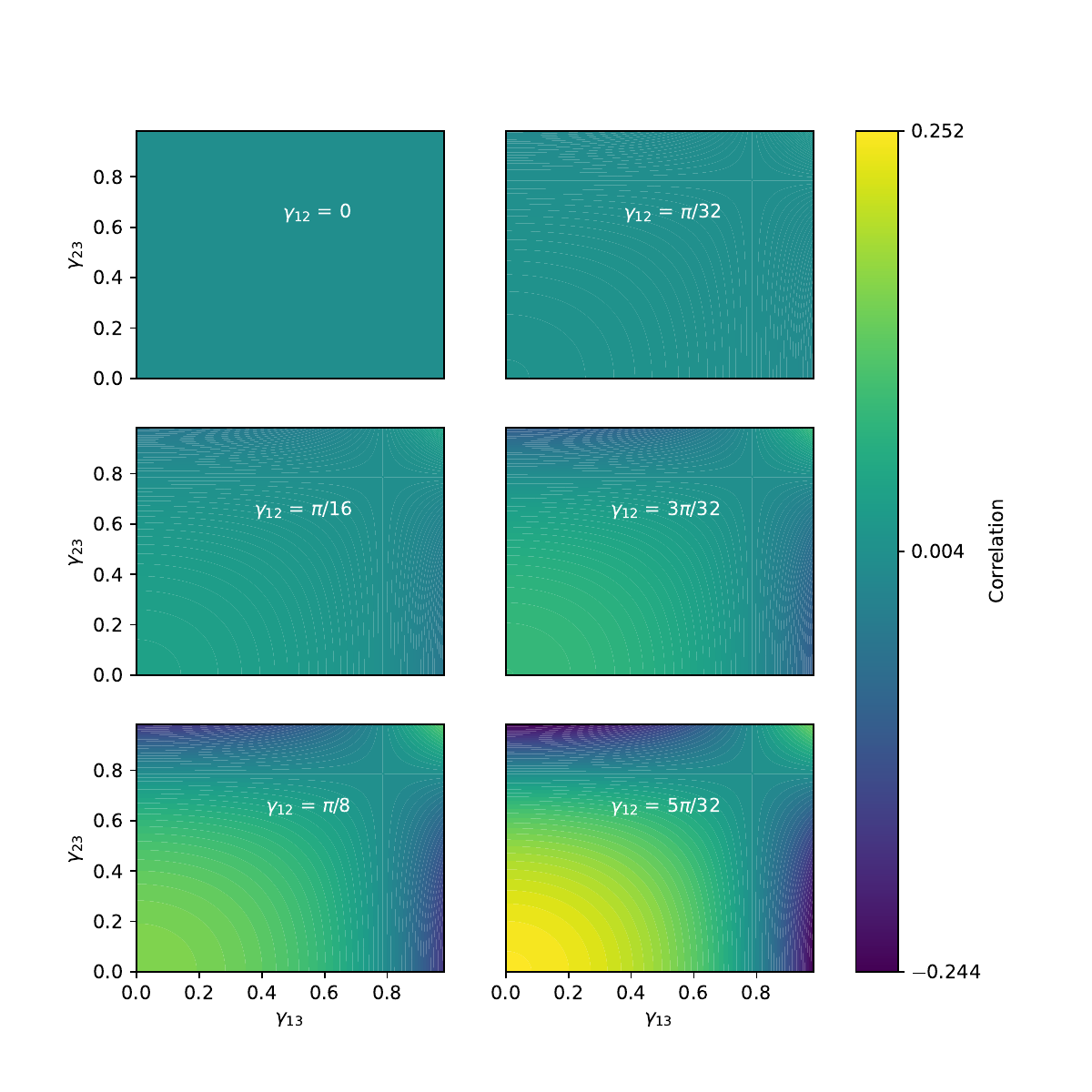}}
\caption{\textbf{These graphs depict the relationship between different coupling strengths $\{\gamma_{12}, \gamma_{13}, \gamma_{23}\}$ and correlation $F_{12}$ within the context of three-gate coupling examples.} Particularly, we assign fixed values to $\gamma_{12}$, selecting from the set $\{0, \pi/32, \pi16, 3\pi/32, \pi/8, 5\pi/32 \}$. The values of $\gamma_{13}$ and $\gamma_{23}$ range from 0 to $5\pi/16$.}
\label{fig:theocorrelation}
\end{figure}

To further investigate how the correlation depends on parameters $\gamma_{12}, \gamma_{13}$, and $\gamma_{23}$, we fix $\gamma_{12}$ with values in $\{0, \pi/32, \pi/16, 3\pi/32, \pi/8, 5\pi/32 \}$ and depict the correlation with respect to $\gamma_{13}$ and $\gamma_{23}$. The results are available in Figure~\ref{fig:theocorrelation}.

From the expressions in Eqs.~\eqref{eq:correlationtwo} and~\eqref{eq:correlationthree}, we observe a natural result that the correlation between two gates always increases with their own coupling strength. In the two-gate coupling case, the correlation is $\sin \gamma_{12}\tan \gamma_{12}$, which increases monotonically with $\gamma_{12}$. In the three-gate coupling case, the numerator of the correlation is proportional to $\cos^2\gamma_{12}\sin^2\gamma_{12}$, which also increases monotonically with $\gamma_{12}$. Thus, the correlation magnitude directly reflects the coupling strength.

Examining the sign of the correlation reveals distinct behaviors. If only two CZ gates are coupled, their correlation is always positive. However, in a three-gate coupling scenario, the situation changes. The correlation is still positive for a low-strength correlated noise when all parameters are less than $\pi/4$. Nonetheless, if two CZ gates exhibit strong $ZZ$ coupling, one of the CZ gates will have a negative correlation with the third CZ gate. From Eq.~\eqref{eq:correlationthree} and Figure~\ref{fig:theocorrelation}, we can see that the negativity becomes particularly pronounced when one coupling is weak while the other is strong. This characteristic is useful for identifying the strongly coupled pair of gates.

Beyond correlation analysis, the dependence of local gate fidelities and global fidelity on the coupling strength helps explain the differences between these two types of fidelities in gate optimization. In the case of $r = 2$, this difference is small. Both kinds of fidelities are proportional to $\cos^2 \gamma_{12}$, and optimization naturally reduces $\gamma_{12}$ to improve performance. Nonetheless, for the three-gate coupling model, the situation can be different. Each local fidelity depends only on two coupling parameters. When optimizing $F_1$, the coupling parameters $\gamma_{12}$ and $\gamma_{13}$ tend to be smaller to make $F_1$ higher and make the correlation weaker. However, given a fixed $\gamma_{23}$, the decrease of $\gamma_{12}$ or $\gamma_{13}$ can also decrease $F_2$ or $F_3$. For instance, $F_2 = p_2(\cos^2 \gamma_{12}(\cos^2 \gamma_{23} - \sin^2 \gamma_{23}) + \sin^2 \gamma_{23}) + \frac{1-p_2}{16}$. When $\cos^2 \gamma_{23} - \sin^2 \gamma_{23} < 0$, the decrease of $\gamma_{12}$ makes $F_2$ also decrease. This creates a competition between optimizing one local fidelity and another, often leading to the optimization being trapped in a local region. In contrast, global fidelity incorporates all coupling parameters, allowing the optimization process to simultaneously reduce all coupling strengths.

%%%%%%%%%%%%%%%%%%%%%%%%%%%%%%%%%%%%%%%%
% choose a style
\bibliographystyle{apsrev4-1}
%%%%%%%%%%%%%%%%%%%%%%%%%%%%%%%%%%%%%%%%
%%%%%%%%%%%%%%%%%%%%%%%%%%%%%%%%%%%%%%%%
% choose a .bib file
% \bibliography{bibPara.bib}

\begin{thebibliography}{99}
\bibitem{aharonov1996limitations}
D. Aharonov, M. Ben-Or, R. Impagliazzo, and N. Nisan, "Limitations of noisy reversible computation (1996), quant - ph/9611028", URL {https://arxiv.org/abs/quant-ph/9611028}

\bibitem{Alexander2016entropy}
A. Müller - Hermes, D. Stilck França, and M. M. Wolf, "Journal of Mathematical Physics 57, 022202 (2016)", ISSN 0022 - 2488, URL {https://doi.org/10.1063/1.4939560}

\bibitem{Daniel2021Limitations}
D. Stilck França and R. García - Patrón, "Nature Physics 17, 1221 (2021)", ISSN 1745 - 2481, URL {https://doi.org/10.1038/s41567 - 021 - 01356 - 3}

\bibitem{Aharonov2023Noisy}
D. Aharonov, X. Gao, Z. Landau, Y. Liu, and U. Vazirani, "in Proceedings of the 55th Annual ACM Symposium on Theory of Computing (Association for Computing Machinery, New York, NY, USA, 2023), STOC 2023", pp. 945 - 957, ISBN 9781450399135, URL {https://doi.org/10.1145/3564246.3585234}

\bibitem{yan2023limitations}
Y. Yan, Z. Du, J. Chen, and X. Ma, "Limitations of noisy quantum devices in computational and entangling power (2023), 2306.02836", URL {https://arxiv.org/abs/2306.02836}

\bibitem{Egan2021corrected}
L. Egan, D. M. Debroy, C. Noel, A. Risinger, D. Zhu, D. Biswas, M. Newman, M. Li, K. R. Brown, M. Cetina, et al., "Nature 598, 281 (2021)", ISSN 1476 - 4687, URL {https://doi.org/10.1038/s41586 - 021 - 03928 - y}

\bibitem{Gong2021correcting}
M. Gong, X. Yuan, S. Wang, Y. Wu, Y. Zhao, C. Zha, S. Li, Z. Zhang, Q. Zhao, Y. Liu, et al., "National Science Review 9, nwab011 (2021)", ISSN 2095 - 5138, URL {https://doi.org/10.1093/nsr/nwab011}

\bibitem{Postler2022tolerant}
L. Postler, S. Heußen, I. Pogorelov, M. Rispler, T. Feldker, M. Meth, C. D. Marciniak, R. Stricker, M. Ringbauer, R. Blatt, et al., "Nature 605, 675 (2022)", ISSN 1476 - 4687, URL {https://doi.org/10.1038/s41586 - 022 - 04721 - 1}

\bibitem{Zhao2022Correcting}
Y. Zhao, Y. Ye, H. - L. Huang, Y. Zhang, D. Wu, H. Guan, Q. Zhu, Z. Wei, T. He, S. Cao, et al., "Phys. Rev. Lett. 129, 030501 (2022)", URL {https://link.aps.org/doi/10.1103/PhysRevLett.129.030501}

\bibitem{Acharya2023Suppressing}
R. Acharya, I. Aleiner, R. Allen, T. I. Andersen, M. Ansmann, F. Arute, K. Arya, A. Asfaw, J. Atalaya, R. Babbush, et al., "Nature 614, 676 (2023)", ISSN 1476 - 4687, URL {https://doi.org/10.1038/s41586 - 022 - 05434 - 1}

\bibitem{Kitaev1997computations}
A. Y. Kitaev, "Russian Mathematical Surveys 52, 1191 (1997)", URL {https://dx.doi.org/10.1070/RM1997v052n06ABEH002155}

\bibitem{aharonov1997fault}
D. Aharonov and M. Ben-Or, "in Proceedings of the Twenty-Ninth Annual ACM Symposium on Theory of Computing (Association for Computing Machinery, New York, NY, USA, 1997), STOC ’97", pp. 176 - 188, ISBN 0897918886, URL {https://doi.org/10.1145/258533.258579}

\bibitem{Knill1998thresholds}
E. Knill, R. Laflamme, and W. H. Zurek, "Proceedings of the Royal Society of London. Series A: Mathematical, Physical and Engineering Sciences 454, 365 (1998)", URL {https://royalsocietypublishing.org/doi/abs/10.1098/rspa.1998.0166}

\bibitem{barends2014logic}
R. Barends, J. Kelly, A. Megrant, A. Veitia, D. Sank, E. Jeffrey, T. C. White, J. Mutus, A. G. Fowler, B. Campbell, et al., "Nature 508, 500 (2014)", URL {https://doi.org/10.1038/nature13171}

\bibitem{Chuang1997tomo}
I. L. Chuang and M. A. Nielsen, "Journal of Modern Optics 44, 2455 (1997)", URL {https://www.tandfonline.com/doi/abs/10.1080/09500349708231894}

\bibitem{Flammia2011prlDirectFidelity}
S. T. Flammia and Y. - K. Liu, "Phys. Rev. Lett. 106, 230501 (2011)", URL {https://link.aps.org/doi/10.1103/PhysRevLett.106.230501}

\bibitem{Emerson2005Scalable}
J. Emerson, R. Alicki, and K. Zyczkowski, "Journal of Optics B: Quantum and Semiclassical Optics 7, S347 (2005)", URL {https://doi.org/10.1088/1464 - 4266/7/10/021}

\bibitem{Emerson2007Characterization}
J. Emerson, M. Silva, O. Moussa, C. Ryan, M. Laforest, J. Baugh, D. G. Cory, and R. Laflamme, "Science 317, 1893 (2007)", URL {https://www.science.org/doi/abs/10.1126/science.1145699}

\bibitem{Knill2008RB}
E. Knill, D. Leibfried, R. Reichle, J. Britton, R. B. Blakestad, J. D. Jost, C. Langer, R. Ozeri, S. Seidelin, and D. J. Wineland, "Phys. Rev. A 77, 012307 (2008)", URL {https://link.aps.org/doi/10.1103/PhysRevA.77.012307}

\bibitem{Emerson2011prlRB}
E. Magesan, J. M. Gambetta, and J. Emerson, "Phys. Rev. Lett. 106, 180504 (2011)", URL {https://link.aps.org/doi/10.1103/PhysRevLett.106.180504}

\bibitem{Emerson2012praRB}
E. Magesan, J. M. Gambetta, and J. Emerson, "Phys. Rev. A 85, 042311 (2012)", URL {https://link.aps.org/doi/10.1103/PhysRevA.85.042311}

\bibitem{Magesan2012interleavedRB}
E. Magesan, J. M. Gambetta, B. R. Johnson, C. A. Ryan, J. M. Chow, S. T. Merkel, M. P. da Silva, G. A. Keefe, M. B. Rothwell, T. A. Ohki, et al., "Phys. Rev. Lett. 109, 080505 (2012)", URL {https://link.aps.org/doi/10.1103/PhysRevLett.109.080505}

\bibitem{McKay2019Three}
D. C. McKay, S. Sheldon, J. A. Smolin, J. M. Chow, and J. M. Gambetta, "Phys. Rev. Lett. 122, 200502 (2019)", URL {https://link.aps.org/doi/10.1103/PhysRevLett.122.200502}

\bibitem{Proctor2019DirectRB}
T. J. Proctor, A. Carignan - Dugas, K. Rudinger, E. Nielsen, R. Blume - Kohout, and K. Young, "Phys. Rev. Lett. 123, 030503 (2019)", URL {https://link.aps.org/doi/10.1103/PhysRevLett.123.030503}

\bibitem{Erhard2019cycleRB}
A. Erhard, J. J. Wallman, L. Postler, M. Meth, R. Stricker, E. A. Martinez, P. Schindler, T. Monz, J. Emerson, and R. Blatt, "Nature Communications 10, 5347 (2019)", ISSN 2041 - 1723, URL {https://doi.org/10.1038/s41467 - 019 - 13068 - 7}

\bibitem{hines2022demonstrating}
J. Hines, M. Lu, R. K. Naik, A. Hashim, J. - L. Ville, B. Mitchell, J. M. Kriekebaum, D. I. Santiago, S. Seritan, E. Nielsen, et al., "Phys. Rev. X 13, 041030 (2023)", URL {https://link.aps.org/doi/10.1103/PhysRevX.13.041030}

\bibitem{zhu2022}
Q. Zhu, S. Cao, F. Chen, M. - C. Chen, X. Chen, T. - H. Chung, H. Deng, Y. Du, D. Fan, M. Gong, et al., "Science Bulletin 67, 240 (2022)", ISSN 2095 - 9273, URL {https://www.sciencedirect.com/science/article/pii/S2095927321006733}

\bibitem{morvan2024phase}
A. Morvan, B. Villalonga, X. Mi, S. Mandrá, A. Bengtsson, P. V. Klimov, Z. Chen, S. Hong, C. Erickson, I. K. Drozdov, et al., "Nature 634, 328 (2024)", ISSN 1476 - 4687, URL {https://doi.org/10.1038/s41586 - 024 - 07998 - 6}

\bibitem{mckay2023benchmarking}
D. C. McKay, I. Hincks, E. J. Pritchett, M. Carroll, L. C. G. Govia, and S. T. Merkel, "Benchmarking quantum processor performance at scale (2023), 2311.05933", URL {https://arxiv.org/abs/2311.05933}

\bibitem{Moses2023Ion}
S. A. Moses, C. H. Baldwin, M. S. Allman, R. Ancona, L. Ascarrunz, C. Barnes, J. Bartolotta, B. Bjork, P. Blanchard, M. Bohn, et al., "Phys. Rev. X 13, 041052 (2023)", URL {https://link.aps.org/doi/10.1103/PhysRevX.13.041052}

\bibitem{Joshi2023large}
M. K. Joshi, C. Kokail, R. van Bijnen, F. Kranzl, T. V. Zache, R. Blatt, C. F. Roos, and P. Zoller, "Nature 624, 539 (2023)", ISSN 1476 - 4687, URL {https://doi.org/10.1038/s41586 - 023 - 06768 - 0}

\bibitem{Bluvstein2024atom}
D. Bluvstein, S. J. Evered, A. A. Geim, S. H. Li, H. Zhou, T. Manovitz, S. Ebadi, M. Cain, M. Kalinowski, D. Hangleiter, et al., "Nature 626, 58 (2024)", URL {https://doi.org/10.1038/s41586 - 023 - 06927 - 3}

\bibitem{Zhang2023cab}
Y. Zhang, W. Yu, P. Zeng, G. Liu, and X. Ma, "Photon. Res. 11, 81 (2023)", URL {https://opg.optica.org/prj/abstract.cfm?URI = prj - 11 - 1 - 81}

\bibitem{Cerezo2021VQA}
M. Cerezo, A. Arrasmith, R. Babbush, S. C. Benjamin, S. Endo, K. Fujii, J. R. McClean, K. Mitarai, X. Yuan, L. Cincio, et al., "Nature Reviews Physics 3, 625 (2021)", URL {https://doi.org/10.1038/s42254 - 021 - 00348 - 9}

\bibitem{Raussendorf2003mbqc}
R. Raussendorf, D. E. Browne, and H. J. Briegel, "Physical Review A 68 (2003)", URL {https://doi.org/10.1103\%2Fphysreva.68.022312}

\bibitem{Hein2004graph}
M. Hein, J. Eisert, and H. J. Briegel, "Phys. Rev. A 69, 062311 (2004)", URL {https://link.aps.org/doi/10.1103/PhysRevA.69.062311}

\bibitem{nelder1965simplex}
J. A. Nelder and R. Mead, "The Computer Journal 7, 308 (1965)", ISSN 0010 - 4620, URL {https://doi.org/10.1093/comjnl/7.4.308}
\bibitem{supplementary}
See Supplementary Material for more detailed experimental data of the comparison between character-average benchmarking and cycle benchmarking, the fluctuation analysis of the experimental results, and additional benchmarking and optimization results, which includes references~\cite{Zhang2023cab,Erhard2019cycleRB}.

\bibitem{rol2017restless}
M. A. Rol, C. C. Bultink, T. E. O'Brien, S. R. de Jong, L. S. Theis, X. Fu, F. Luthi, R. F. L. Vermeulen, J. C. de Sterke, A. Bruno, et al., ``Phys. Rev. Appl. 7, 041001 (2017)'', URL {https://link.aps.org/doi/10.1103/PhysRevApplied.7.041001}

\bibitem{gottesman1997stabilizer}
D. Gottesman, ``Stabilizer codes and quantum error correction (California Institute of Technology, 1997)'', URL {https://arxiv.org/abs/quant - ph/9705052}

\bibitem{Bravyi2024Highthreshold}
A. W. Cross, J. M. Gambetta, D. Maslov, P. Rall, and T. J. Yoder, ``Nature 627, 778 (2024)'', URL {https://doi.org/10.1038/s41586-024-07107-7}

\bibitem{YulinWu2021Superconducting}
Y. Wu, W. - S. Bao, S. Cao, F. Chen, M. - C. Chen, X. Chen, T. - H. Chung, H. Deng, Y. Du, D. Fan, et al., ``Phys. Rev. Lett. 127, 180501 (2021)'', URL {https://link.aps.org/doi/10.1103/PhysRevLett.127.180501}

\bibitem{Barends0Diabatic}
R. Barends, C. M. Quintana, A. G. Petukhov, Y. Chen, D. Kafri, K. Kechedzhi, R. Collins, O. Naaman, S. Boixo, F. Arute, et al., ``Phys. Rev. Lett. 123, 210501 (2021)'', URL {https://link.aps.org/doi/10.1103/PhysRevLett.123.210501}

\bibitem{2017Characterizing}
S. Boixo, in ``APS March Meeting Abstracts (2017), vol. 2017 of APS Meeting Abstracts, p. A19.005

\bibitem{Arute2019}
F. Arute, K. Arya, R. Babbrusch, D. Bacon, J. C. Bardin, R. Barends, R. Biswas, S. Boixo, F. G. S. L. Brandao, D. A. Buell, et al., ``Nature 574, 505 (2019)'', ISSN 1476 - 4687, URL {https://doi.org/10.1038/s41586 - 019 - 1666 - 5}

\bibitem{2021Realization}
H. Xu, W. Liu, Z. Li, J. Han, J. Zhang, K. Linghu, Y. Li, M. Chen, Z. Yang, J. Wang, et al., ``Chinese Physics B 30, 044212 (2021)'', URL {https://dx.doi.org/10.1088/1674 - 1056/abf03a}

\end{thebibliography}

\end{document}

\title{Supplemental Material: Calibrating Quantum Gates up to 52 Qubits in a Superconducting Processor}

\author{Daojin Fan}\thanks{These authors contributed equally to this work.}
\affiliation{\adda}
\affiliation{\addb}
\affiliation{\addc}
\author{Guoding Liu}\thanks{These authors contributed equally to this work.}
\affiliation{\addd}
\author{Shaowei Li}\thanks{These authors contributed equally to this work.}
\affiliation{\adda}
\affiliation{\addb}
\affiliation{\addc}
\author{Ming Gong}
\affiliation{\adda}
\affiliation{\addb}
\affiliation{\addc}
\affiliation{\addf}
\author{Dachao Wu}
\affiliation{\adda}
\affiliation{\addb}
\affiliation{\addc}
\author{Yiming Zhang}
\affiliation{\adda}
\affiliation{\addb}
\affiliation{\addc}
\author{Chen Zha}
\affiliation{\adda}
\affiliation{\addb}
\affiliation{\addc}
\author{Fusheng Chen}
\affiliation{\adda}
\affiliation{\addb}
\affiliation{\addc}
\author{Sirui Cao}
\affiliation{\adda}
\affiliation{\addb}
\affiliation{\addc}
\author{Yangsen Ye}
\affiliation{\adda}
\affiliation{\addb}
\affiliation{\addc}
\author{Qingling Zhu}
\affiliation{\adda}
\affiliation{\addb}
\affiliation{\addc}
\author{Chong Ying}
\affiliation{\adda}
\affiliation{\addb}
\affiliation{\addc}
\author{Shaojun Guo}
\affiliation{\adda}
\affiliation{\addb}
\affiliation{\addc}
\author{Haoran Qian}
\affiliation{\adda}
\affiliation{\addb}
\affiliation{\addc}
\author{Yulin Wu} 
\affiliation{\adda}
\affiliation{\addb}
\affiliation{\addc}
\author{Hui Deng} 
\affiliation{\adda}
\affiliation{\addb}
\affiliation{\addc}
\affiliation{\addf}
\author{Gang Wu}
\affiliation{\adda}
\affiliation{\addb}
\affiliation{\addc}
\affiliation{\addf}
\affiliation{\adde}
\author{Cheng-Zhi Peng}
\affiliation{\adda}
\affiliation{\addb}
\affiliation{\addc}
\affiliation{\addf}
\author{Xiongfeng Ma}
\thanks{xma@tsinghua.edu.cn}
\affiliation{\addd}
\author{Xiaobo Zhu}
\thanks{xbzhu16@ustc.edu.cn}
\affiliation{\adda}
\affiliation{\addb}
\affiliation{\addc}
\affiliation{\addf}
\author{Jian-Wei Pan}
\thanks{pan@ustc.edu.cn}
\affiliation{\adda}
\affiliation{\addb}
\affiliation{\addc}
\affiliation{\addf}

\newcommand{\adda}{Hefei National Research Center for Physical Sciences at the Microscale and School of Physical Sciences, University of Science and Technology of China, Hefei 230026, China}
\newcommand{\addb}{Shanghai Research Center for Quantum Science and CAS Center for Excellence in Quantum Information and Quantum Physics, University of Science and Technology of China, Shanghai 201315, China}
\newcommand{\addc}{Hefei National Laboratory, University of Science and Technology of China, Hefei 230088, China}
\newcommand{\addd}{Center for Quantum Information, Institute for Interdisciplinary Information Sciences, Tsinghua University, Beijing, 100084 China}
\newcommand{\adde}{University of Science and Technology of China, Shanghai Research Institute, Shanghai 201315, China}
\newcommand{\addf}{Jinan Institute of Quantum Technology and Hefei National Laboratory Jinan Branch, Jinan 250101,China}

\begin{abstract}
Here, we briefly introduce the content of the Supplemental Material. In Section~\ref{appendsc:comparison}, we present the comparison between character-average benchmarking and cycle benchmarking to demonstrate the soundness and advantage of character-average benchmarking. In Section~\ref{appendsc:add}, we show the fluctuation analysis of the correlation evaluation and additional benchmarking and optimization results with detailed data. In this work, the fluctuation analysis is done by two kinds of methods: one is performing multiple times of experiments and evaluating their standard deviations, which contain fluctuation effects from one experiment and deviations among multiple experiments. The associated results are denoted as SD. The other is done by standard error propagation, which represents the fluctuation effect for a single experiment. The associated results are denoted as SE. We will clarify the detailed procedure for each evaluation.
\end{abstract}

\maketitle

\section{Comparison of character-average benchmarking and cycle benchmarking}\label{appendsc:comparison}
We validate the soundness of character-average benchmarking (CAB)~\cite{Zhang2023cab} by comparing its benchmarking results of the CZ gate to cycle benchmarking (CB)~\cite{Erhard2019cycleRB}. The reason why CB was chosen for comparison is that CAB and CB both benchmark the fidelity of an individual target gate dressed with local Pauli gates. Other RB protocols are normally aimed at benchmarking the fidelity of a gate set, differing from CAB and CB. In fact, CAB is developed from a simple variant of CB. Setting the target gate as the CZ gate instead of other gates is because CZ is easy to realize and has a low order. The property of low order is essential for CB as, within this benchmarking method, the circuit depth is proportional to the order of the target gate. Beyond the soundness validation, we experimentally found that CAB enjoys a smaller variance than CB, enabling us to get the fidelity of a quantum gate in a shorter time without compromising accuracy. Meanwhile, we show two results for CAB, which are associated with two different postprocessing methods. The original one in Ref.~\cite{Zhang2023cab} requires measuring observables traversing $\{\mathbb{I},Z\}^{\otimes n}$, denoted as `CAB, traverse'. The protocol in this work only requires measuring a constant number of observables in $\{\mathbb{I}, Z\}^{\otimes n}$, denoted as `CAB, sample'.

\begin{figure}[!htbp]
\centering \resizebox{17cm}{!}{\includegraphics{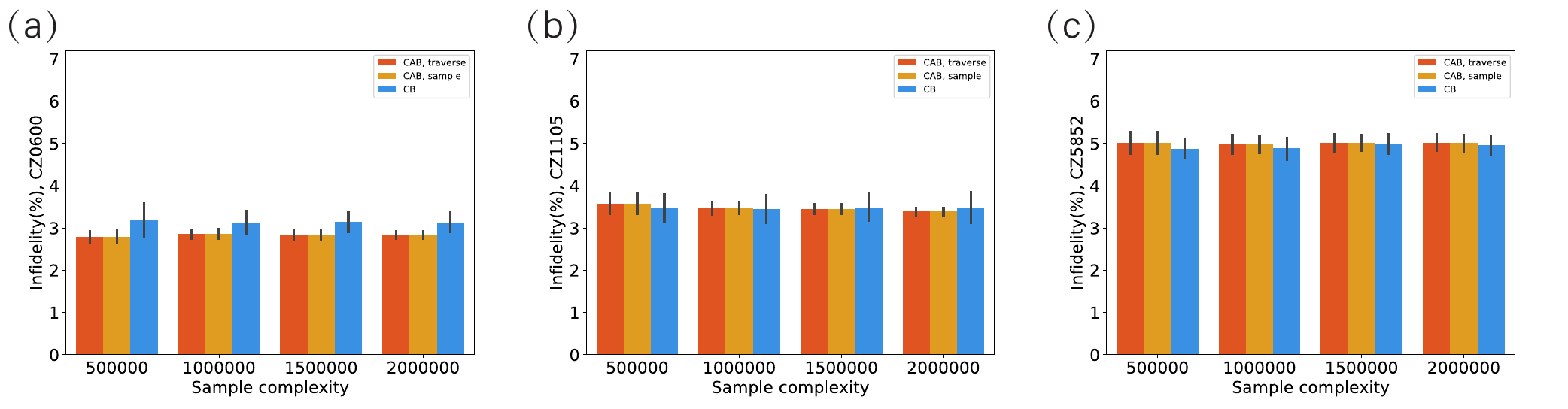}}
\caption{The figures demonstrate the CAB fidelity and CB fidelity comparison for benchmarking the CZ gate. Figures (a), (b), and (c) correspond to three different CZ gates. The fidelities are both the CZ fidelity dressed with the local Pauli gates. The standard deviation, shown with the error bar, is obtained via repeating experiments 25 times and calculating the standard deviation of 25 fidelities. The quantity of `infidelity' is defined as `1-fidelity'. The red and orange bars are both obtained from CAB experimental data but with two different classical postprocessing methods. The former needs estimating the observable value traversing $\{\mathbb{I}, Z\}^{\otimes n}$, and the latter only estimates a constant number of observables from $\{\mathbb{I}, Z\}^{\otimes n}$. The blue bar is obtained from CB experimental data.}
\label{fig:cabvscb}
\end{figure}

In the comparison experiment, we set the list of circuit depths to be $\{5, 10\}$, corresponding to a total number of CZ gates as $\{10, 20\}$, for all benchmarking methods. For each circuit depth, we sample $K_r$ random sequences, and for each sequence, we measure $K_s=20$,$000$ times. The total sample complexity is $K_rK_s$, ignoring the number of circuit depths. Note that in CB, one needs to sample Pauli operators additionally as the character gates. Thus, in CB, the $K_r$ random sequences are divided into 5 groups, each associated with a randomly sampled Pauli operator. The total sample complexity is the same for CAB and CB. The results are shown in Figure~\ref{fig:cabvscb}.

In Figs.~\ref{fig:cabvscb}(a), (b), and (c), we compare the benchmarking results for three fixed CZ gates with respect to the sample complexity ranging from 500,000 to 2,000,000. For `CAB, sample', the number of sampled operators $K_q$ is set as $1000$. From the results, we can see nearly no difference between `CAB, traverse' and `CAB, sample', indicating that sampling observables instead of measuring all of them in $\{\mathbb{I}, Z\}^{\otimes n}$ still allow getting accurate CAB fidelities. Most importantly, the fidelities benchmarked with CAB and CB differ by less than one standard deviation for all three gates, showing the soundness of CAB.

Regarding the fluctuation of the benchmarking results, the standard deviation of CAB is smaller than that of CB. It is reasonable because, in classical processing, CAB utilizes all the information of the measurement results, while CB only cares about part of the information. The advantage of lower fluctuation is essential for gate benchmarking, as in experiments, one needs to iterate the benchmarking and optimization process plenty of times in a limited time. If the fluctuation of the benchmarking process is too large, the number of sampled sequences also needs to be large, and hence, the gate optimization would be slow. The benchmarking results may also be unreliable as the parameters of the pulses to implement quantum gates may drift after a long time.

Note that the standard deviation shown in Figure~\ref{fig:cabvscb} is obtained by implementing 25 different experiments and calculating the standard deviation of the fidelities from different experiments. This standard deviation value contains two effects: one is the self-fluctuation of each experiment, and the other is the difference among different experiments. In practice, we normally implement a single benchmarking experiment and do not repeat it. In this case, the self-fluctuation of a single experiment is important to represent the stability of the benchmarking result. To provide a more comprehensive fluctuation evaluation for the benchmarking methods, we also evaluate the standard error using data from one experiment via the propagation of the standard error. Below, we briefly sketch this method for CAB, and the procedure is similar for CB.

In experiments, the directly available data is the frequency of the measured bitstrings in $\{0,1\}^n$. Viewing it as a multinomial distribution, we can get the standard error of each measured bitstring. Since the survival probability is a linear combination of the frequencies of the measured bitstrings, we can evaluate the standard error of the survival probability for each depth and each observable and, more concretely, the covariance matrix of different survival probabilities. Then, via the propagation of the standard error of fitting, we can get the standard error of each noise channel quality parameter and the covariance matrix among different quality parameters. Based on that, the final fidelity is the average of the noise channel quality parameters, we can evaluate the standard error of the fidelity. It is worth mentioning that when the number of survival probabilities is two, the fitting can be done perfectly; there will be no fitting error in this case, and the standard error will be small. This is also why we use data from two depths to evaluate the fidelity during our experiments.

The fidelity, standard deviation, and standard error data are shown in Table~\ref{tab:cabvscb}. It can be seen that regardless of standard deviation or standard error, the fluctuation of CB is always larger than CAB. The results indicate a lower fluctuation of CAB. It is worth mentioning that the standard deviation is larger than the standard error. This difference comes from the deviation among different experiments. As the pulse parameters for implementing quantum gates can drift over a long time, the difference among different experiments can be large and dominate the standard deviation value.

\begin{table}[!htbp]
\centering
\caption{The table shows the fidelity estimation along with the standard deviation of the estimation for three benchmarking procedures and three CZ gates corresponding to Figure~\ref{fig:cabvscb}. We also show the average standard error of the fidelity estimation over 25 experiments. The standard deviation, denoted by SD, is evaluated by implementing multiple experiments and calculating the standard deviation of 25 estimated fidelities. The standard error, denoted by SE, is calculated using one experiment data via the standard error propagation formula. Note that we can get 25 different standard error values, and here, SE represents the average of these 25 values.}

\begin{tabular}{cccccccccc}\hline
CZ0600 & \multicolumn{3}{c}{CAB, traverse} & \multicolumn{3}{c}{CAB, sample} & \multicolumn{3}{c}{CB} \\\hline
Sample complexity & Fidelity & SD & SE & Fidelity & SD & SE & Fidelity & SD & SE \\\hline
500,000 & 97.21\% & 0.38\% & 0.02\% & 97.21\% & 0.39\% & 0.02\% & 96.82\% & 1.02\% & 0.03\% \\
1,000,000 & 97.14\% & 0.27\% & 0.02\% & 97.14\% & 0.27\% & 0.02\% & 96.87\% & 0.73\% & 0.02\% \\
1,500,000 & 97.16\% & 0.27\% & 0.01\% & 97.16\% & 0.26\% & 0.01\% & 96.86\% & 0.63\% & 0.02\% \\
2,000,000 & 97.16\% & 0.23\% & 0.01\% & 97.17\% & 0.23\% & 0.01\% & 96.87\% & 0.58\% & 0.02\% \\\hline
\end{tabular}

\begin{tabular}{cccccccccc}\hline
CZ1105 & \multicolumn{3}{c}{CAB, traverse} & \multicolumn{3}{c}{CAB, sample} & \multicolumn{3}{c}{CB} \\\hline
Sample complexity & Fidelity & SD & SE & Fidelity & SD & SE & Fidelity & SD & SE \\\hline
500,000 & 96.42\% & 0.66\% & 0.02\% & 96.42\% & 0.65\% & 0.02\% & 96.54\% & 0.89\% & 0.03\% \\
1,000,000 & 96.53\% & 0.38\% & 0.02\% & 96.53\% & 0.38\% & 0.02\% & 96.55\% & 0.86\% & 0.02\% \\
1,500,000 & 96.56\% & 0.29\% & 0.01\% & 96.55\% & 0.31\% & 0.01\% & 96.53\% & 0.83\% & 0.02\% \\
2,000,000 & 96.61\% & 0.23\% & 0.01\% & 96.61\% & 0.23\% & 0.01\% & 96.53\% & 0.90\% & 0.01\% \\\hline
\end{tabular}

\begin{tabular}{cccccccccc}\hline
CZ5852 & \multicolumn{3}{c}{CAB, traverse} & \multicolumn{3}{c}{CAB, sample} & \multicolumn{3}{c}{CB} \\\hline
Sample complexity & Fidelity & SD & SE & Fidelity & SD & SE & Fidelity & SD & SE \\\hline
500,000 & 94.99\% & 0.68\% & 0.03\% & 94.98\% & 0.69\% & 0.03\% & 95.12\% & 0.63\% & 0.04\% \\
1,000,000 & 95.02\% & 0.57\% & 0.02\% & 95.03\% & 0.56\% & 0.02\% & 95.11\% & 0.66\% & 0.03\% \\
1,500,000 & 94.99\% & 0.49\% & 0.02\% & 94.99\% & 0.50\% & 0.02\% & 95.05\% & 0.59\% & 0.02\% \\
2,000,000 & 94.98\% & 0.50\% & 0.02\% & 94.99\% & 0.51\% & 0.02\% & 95.04\% & 0.60\% & 0.02\% \\\hline
\end{tabular}

\label{tab:cabvscb}
\end{table}

\section{Additional benchmarking results}\label{appendsc:add}
In this part, we present additional experimental results, including detailed fidelity data for the fully connected gate and the parallel CZ gate, the fluctuation analysis of the correlation estimation, the crosstalk analysis within the parallel CZ gate, and additional optimization results.

\subsection{Detailed benchmarking results for the fully connected quantum gate}\label{appendssc:fully}
Below, we show the detailed benchmarking fidelities along with their standard error for the fully connected gate in Table~\ref{tab:vqe_fidelity}. The standard error is evaluated by the propagation of the standard error.
\begin{table}[!htbp]
\centering
\caption{The fidelity and standard error data of fully connected quantum gates. The standard error is denoted as SE.}
\begin{tabular}{ccccccccc}\hline
Qubit number & 16 & 18 & 20 & 22 & 24 & 26 & 28 & 30 \\\hline
Fidelity & 63.49\% & 58.26\% & 54.88\% & 48.11\% & 47.78\% & 43.04\% & 38.60\% & 37.32\% \\
SE & 0.07\% & 0.08\% & 0.08\% & 0.09\% & 0.09\% & 0.10\% & 0.11\% & 0.16\% \\\hline
Qubit number & 32 & 34 & 36 & 38 & 40 & 42 & 44 & 46 \\\hline
Fidelity & 34.69\% & 31.46\% & 30.11\% & 25.41\% & 22.45\% & 22.37\% & 20.24\% & 17.42\% \\
SE & 0.25\% & 0.29\% & 0.30\% & 0.32\% & 0.35\% & 0.37\% & 0.37\% & 0.45\% \\\hline
\end{tabular}
\label{tab:vqe_fidelity}
\end{table}

\begin{figure}[!htbp]
\centering \resizebox{7cm}{!}{\includegraphics{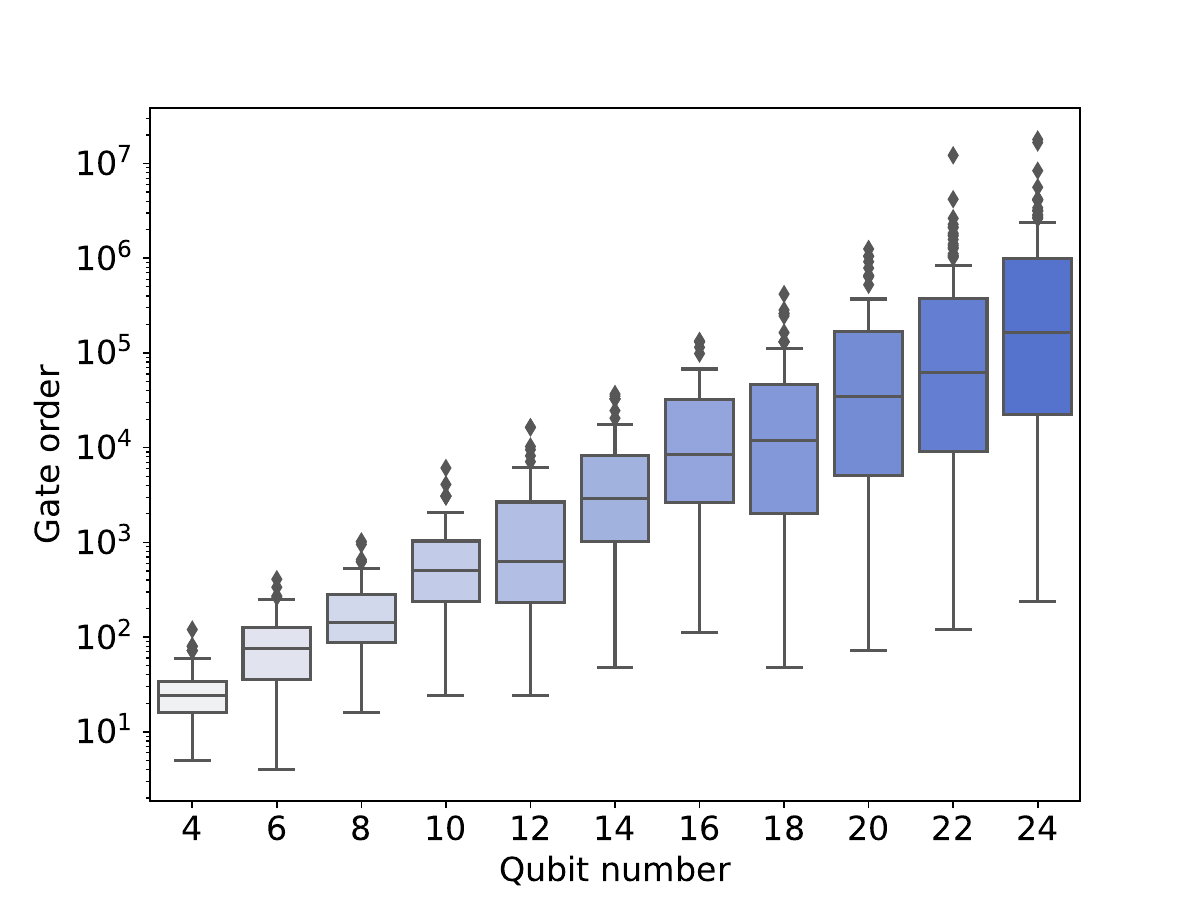}}
\caption{Circuit order distribution for the fully connected gate with qubit number from 4 to 24. For each qubit number, we sample two layers of local Clifford gates 100 times and obtain 100 different gate orders. The figure demonstrates the distribution of 100 orders with box plots. The middle line is the median. The box contains data from the quartile to the three-quarter quartile.}
\label{fig:order}
\end{figure}

In the main text, we mentioned that the fully connected gate generally has a large gate order, so it is challenging to benchmark such a gate via CB. Below, we present a simulation result showing the gate order distribution of the fully connected gate in Figure~\ref{fig:order}. Note that the fully connected gate comprises two layers of randomly sampled local Clifford gates. Different choices of local Clifford gates will influence the gate order. In the simulation, we sample two layers of local Clifford gates 100 times, and each time, we calculate the gate order. We present the 100 gate orders with box plots. This procedure is progressively done from 4 qubits to 24 qubits. The results show that, on average, the gate order of the fully connected gate increases exponentially with respect to the qubit number. The order for 16 qubits approximately ranges from 100 to 10,000 and is thousands on average. On the other hand, the fidelity of the 16-qubit fully connected gate is $63.49\%\pm 0.07\%$. Repeatedly implementing the target gate 100 times will lead to a survival probability of about $10^{-20}$, which is much less than the device accuracy. In this sense, it is hard to use CB to benchmark the fully connected gate with randomly sampled local Clifford gates.

\subsection{Parallel CZ gate fidelity}\label{appendssc:paracz}
Below, we show the fidelities of the parallel CZ gates along with their standard error and summarize in Table~\ref{tab:cz_fidelity}. The standard error of the estimated fidelity is evaluated by the propagation of the standard error. We first calculate the standard error of the measurement results and then calculate that of the survival probability. After that, we get the standard error of the dressed CZ fidelities and local gate fidelities. As the pure CZ fidelity is evaluated by
\begin{equation}
F = \frac{4^nF_{dress}-1}{4^nF_{twirl}-1}(1-\frac{1}{4^n})+\frac{1}{4^n},
\end{equation}
where $F_{dress}$ and $F_{twirl}$ are the dressed and twirling gate fidelities, respectively; $n$ is the qubit number. The standard error of the pure CZ fidelity is estimated by 
\begin{equation}
\sigma(F) = (F-\frac{1}{4^n})\sqrt{\frac{\sigma^2(F_{dress})}{(F_{dress}-\frac{1}{4^n})^2}+\frac{\sigma^2(F_{twirl})}{(F_{twirl}-\frac{1}{4^n})^2}},
\end{equation}
where $\sigma(F)$ means the standard error of $F$.

\begin{table}[!htbp]
\centering
\caption{Fidelity and standard error estimations of the parallel CZ gate within the orange pattern in the superconducting processor and of the 26-pair parallel CZ gate. The standard error is estimated by the propagation of the standard error. The first table shows the results of the CZ gates within the orange pattern, whose fidelities are evaluated with measurement data from depths $\{0, 2\}$. The second table shows the results of the 26-pair parallel CZ gate. The fidelity of the dressed gate is evaluated with measurement data from depths $\{0, 1\}$, and that of the local twirling gates is evaluated with measurement data from depths $\{1, 2\}$. The number of sampled sequences is $K_r = 50$, and the number of measured times for each sequence is $K_s = 20$,$000$. Due to the lower fidelities and closer circuit depths, the standard error of the results of the 26-pair parallel CZ gate is much larger than that of other parallel CZ gates.}

\begin{tabular}{ccccccc}\hline
Orange pattern & \multicolumn{2}{c}{Dressed CZ} & \multicolumn{2}{c}{Local gate} & \multicolumn{2}{c}{Pure CZ} \\\hline
Qubit number & Fidelity & SE & Fidelity & SE & Fidelity & SE \\\hline
4 & 95.48\% & 0.02\% & 98.97\% & 0.02\% & 96.47\% & 0.02\% \\  
8 & 89.89\% & 0.03\% & 97.60\% & 0.02\% & 92.09\% & 0.04\% \\  
12 & 83.96\% & 0.04\% & 95.08\% & 0.03\% & 88.31\% & 0.05\% \\  
16 & 79.57\% & 0.05\% & 94.33\% & 0.04\% & 84.35\% & 0.06\% \\  
20 & 75.43\% & 0.07\% & 92.37\% & 0.05\% & 81.67\% & 0.08\% \\  
24 & 69.93\% & 0.08\% & 90.98\% & 0.05\% & 76.86\% & 0.10\% \\  
28 & 66.74\% & 0.09\% & 89.95\% & 0.06\% & 74.20\% & 0.11\% \\  
32 & 62.10\% & 0.11\% & 87.73\% & 0.06\% & 70.79\% & 0.14\% \\  
36 & 59.97\% & 0.13\% & 86.84\% & 0.07\% & 69.06\% & 0.16\% \\  
40 & 54.91\% & 0.16\% & 85.46\% & 0.08\% & 64.26\% & 0.20\% \\  
44 & 52.74\% & 0.18\% & 83.60\% & 0.09\% & 63.09\% & 0.23\% \\\hline
\end{tabular}

\begin{tabular}{ccccccc}\hline
Blue pattern & \multicolumn{2}{c}{Dressed CZ} & \multicolumn{2}{c}{Local gate} & \multicolumn{2}{c}{Pure CZ} \\\hline
Qubit number & Fidelity & SE & Fidelity & SE & Fidelity & SE \\\hline
52 & 31.43\% & 0.51\% & 80.75\% & 0.87\% & 38.92\% & 0.76\% \\\hline
\end{tabular}

\label{tab:cz_fidelity}
\end{table}

\subsection{Fluctuation of the correlation estimation}
\begin{figure*}[!htbp]
\centering \resizebox{17cm}{!}{\includegraphics{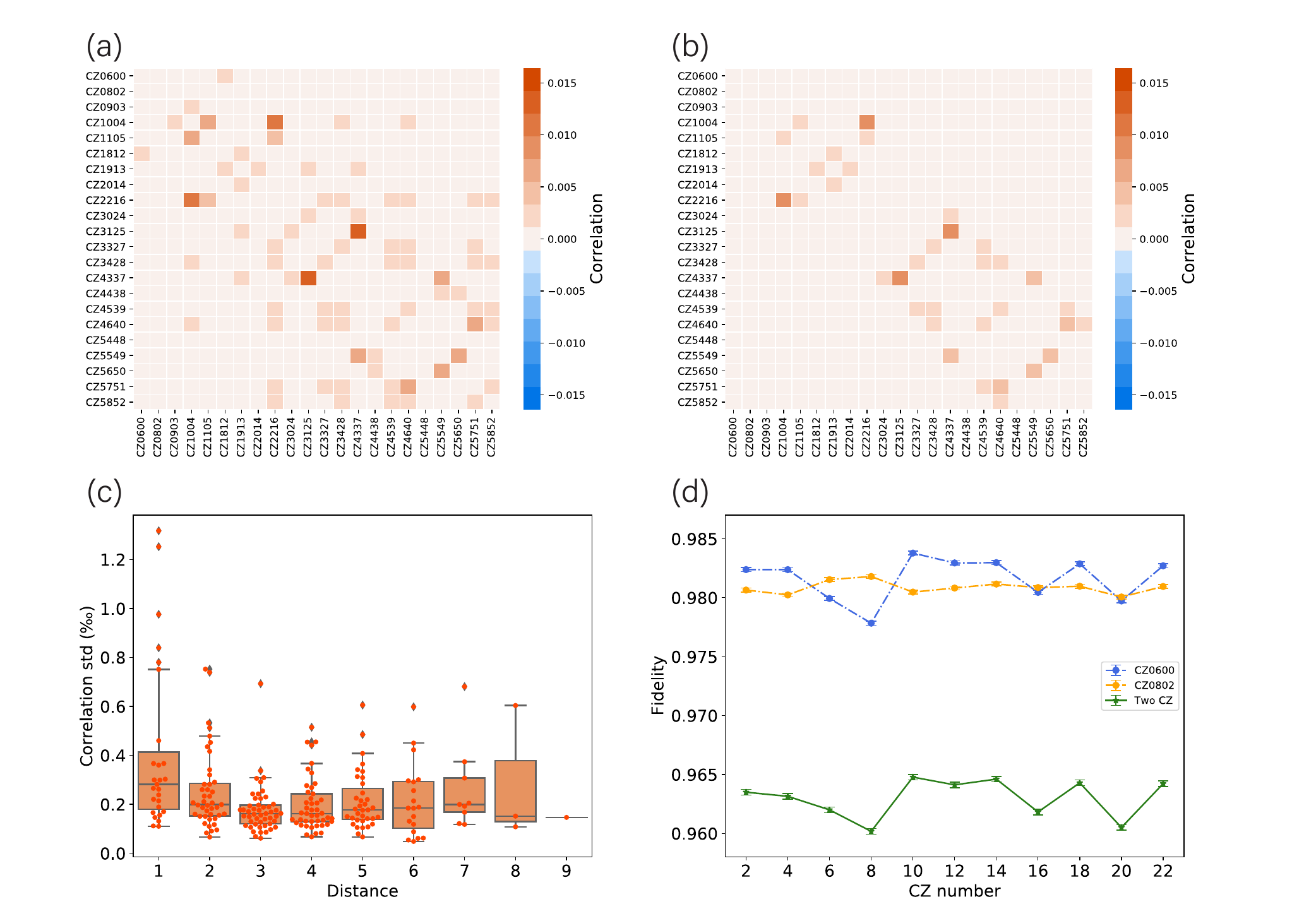}}
\caption{Correlation and crosstalk. (a) The mean value of the 2-correlation among the 22-pair parallel CZ gate. (b) The lower bound of the absolute value of the 2-correlation among the 22-pair parallel CZ gate. It is calculated by $\max(\abs{C_{mean}}-3C_{std}, 0)$ where $C_{mean}$ and $C_{std}$ are the mean value and the standard deviation of the correlation shown in (a) and (c), respectively. (c) The standard deviation of the 2-correlation among the 22-pair parallel CZ gate with respect to the distance between CZ gates. (d) The fidelities of the first CZ, the second CZ, and the first two CZ gates when there are 2, 4, ..., 22 CZ gates in implementation. CZ0600 is the first CZ gate, and CZ0802 is the second CZ gate. The standard error is evaluated by the propagation of the standard error. The data is the same as the one providing the overall fidelity of the parallel CZ gate within the orange pattern.}
\label{fig:fig7}
\end{figure*}

In the main text, we show the correlation between every two gates in the 22-pair parallel CZ gate and 26-pair parallel CZ gate, respectively. There, we only show the value of the correlation, ignoring its fluctuation. To show that the correlation value indeed indicates an interaction between different local gates instead of just fluctuating around zero, we performed an experiment to obtain the level of the fluctuation of the correlation. Specifically, we performed an experiment within the orange pattern, using 22 CZ gates and 44 qubits. We repeat the CAB protocol on these qubits 10 times where the circuit depths are $\{0, 2\}$, the number of sampled sequences for each depth is 50, and the measured times for each sequence is 20,000. Each time, we get the fidelities of the dressed CZ gates and the local twirling gates and use the two fidelities to calculate the pure CZ fidelity. The correlation is evaluated only using the individual fidelity of the parallel CZ gates, excluding the noise effect from the twirling gates. We calculate the mean value and the standard deviation of the correlation values, shown in Figs.~\ref{fig:fig7}(a) and~\ref{fig:fig7}(c), respectively.

It can be seen that the level of the standard deviation is mostly smaller than $10^{-3}$. Thus, we can ensure that if the correlation value is about $10^{-2}$, then it definitely indicates a strong correlation between two gates instead of being a fluctuation. Moreover, from Figure~\ref{fig:fig7}(c), we can see that the value of the standard deviation is nearly irrelevant to the distance between CZ gates. In Figure~\ref{fig:fig7}(b), we also show the lower bound of the absolute value of the correlation, considering a confidence interval three times the standard deviation away from the mean value. The grids in orange indicate an interaction between two gates with high probability.

\subsection{Crosstalk}
In the main text, we have shown that the fidelity of a single CZ gate within the orange pattern does not change much when the number of CZ gates increases. Here, we show an additional figure as a supplement to this result. In implementing the parallel CZ gates using 2, 4, ..., 22 gates within the orange pattern, the parallel CZ gates all contain two same CZ gates -- CZ0600 and CZ0802. We plot their individual fidelities and the fidelity of these two CZ gates in Figure~\ref{fig:fig7}(d) with detailed data in Table~\ref{tab:crosstalk} when implementing the parallel CZ gates using 2, 4, ..., 22 gates. With the total number of CZ gates increases, the fluctuation of the fidelities, which is calculated by the relative difference of the minimum and maximum fidelities, does not overpass 0.61\%, 0.18\%, and 0.48\% for CZ0600, CZ0802, and two CZ gates, respectively. This means that implementing additional CZ gates does not influence much on these two CZ gates. It is in accordance with the result that the fidelity of a single CZ gate does not change much and indicates a weak correlation within the orange pattern.

\begin{table}
\caption{The fidelity and standard error data of crosstalk associated with Figure~\ref{fig:fig7}(d).}
\begin{tabular}{ccccccccccccc}  
\hline 
CZ index & Qubit number & 4 & 8 & 12 & 16 & 20 & 24 & 28 & 32 & 36 & 40 & 44 \\\hline  
\multirow{2}{*}{CZ0600} & Fidelity & 98.24\% & 98.24\% & 97.99\% & 97.78\% & 98.38\% & 98.29\% & 98.30\% & 98.05\% & 98.29\% & 97.97\% & 98.27\% \\
 & SE & 0.02\% & 0.02\% & 0.02\% & 0.02\% & 0.02\% & 0.02\% & 0.02\% & 0.02\% & 0.02\% & 0.02\% & 0.02\% \\ \hline  
\multirow{2}{*}{CZ0802} & Fidelity & 98.06\% & 98.02\% & 98.15\% & 98.18\% & 98.05\% & 98.08\% & 98.12\% & 98.09\% & 98.10\% & 98.01\% & 98.10\% \\ 
 & SE & 0.02\% & 0.02\% & 0.02\% & 0.02\% & 0.02\% & 0.02\% & 0.02\% & 0.02\% & 0.02\% & 0.02\% & 0.02\% \\ \hline  
\multirow{2}{*}{TwoCZ} & Fidelity & 96.35\% & 96.31\% & 96.20\% & 96.02\% & 96.48\% & 96.41\% & 96.46\% & 96.18\% & 96.43\% & 96.05\% & 96.42\% \\
 & SE & 0.02\% & 0.02\% & 0.02\% & 0.02\% & 0.02\% & 0.02\% & 0.02\% & 0.02\% & 0.02\% & 0.02\% & 0.02\% \\ \hline  
\end{tabular}
\label{tab:crosstalk}
\end{table}

\subsection{Additional optimization results}\label{appendssc:opt}
In this part, we present the concrete optimization data for the 3-pair parallel CZ gate as a complement to the optimization results in the main text and show an additional optimization experiment result for a parallel CZ gate with 2 local CZ gates.

\begin{figure}[!htbp]
\centering \resizebox{18cm}{!}{\includegraphics{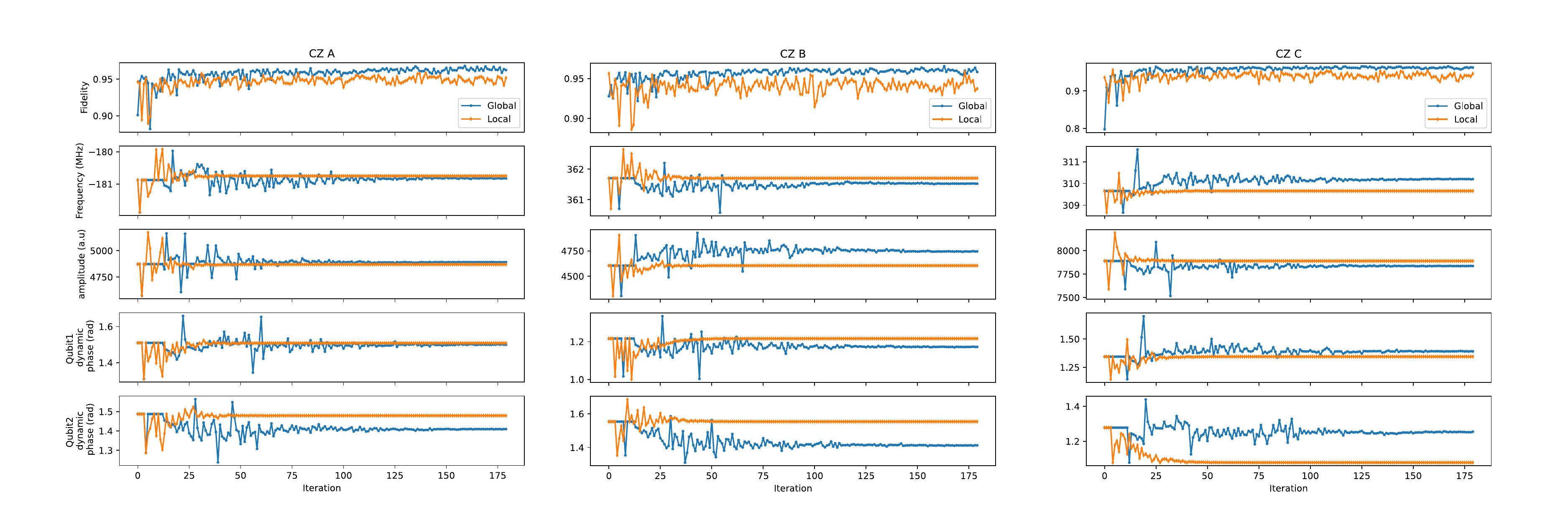}}
\caption{The data of local CZ gate fidelities and gate parameters during the optimization procedure. The dotted lines in blue and orange respectively represent the results obtained by taking the global fidelity and the local fidelity as the target function.}
\label{fig:u_opt}
\end{figure}

We first provide the data for the 3-pair parallel CZ gate. Figure~\ref{fig:u_opt} illustrates the variation in gate parameters and local CZ gate fidelities throughout the optimization process. It should be noted that the data of global fidelity has already been provided in the main text. During the experiment, when optimizing with the global fidelity as the target function, the three sets of gate parameters were integrated into a single parameter space. This method led to a relatively slow convergence rate, yet it produced a higher fidelity upon convergence. From the figure, it can be seen that the gate parameters and fidelities converge and remain relatively stable after approximately 100 iteration steps. Therefore, we can infer that the optimization attains an optimal value, at least within a local area. The statistics of the data for the 3-pair parallel CZ gate are in Table~\ref{tab:opt3CZ}. The results show minimal improvement when using individual local CZ gate fidelities. The fidelity of CZ A is invariant, the fidelity of CZ B decreases a little, and the fidelity of CZ C increases. This can be explained by the fact that local gate fidelities do not contain the information of gate correlation. The optimization of one gate may negatively influence the performance of another gate. Note that the local CZ gates have already been roughly calibrated via fast calibration and with back probability. The local fidelity values may be in a local optimum point. It is hard to optimize the gates further if only local gate fidelities are used.

Conversely, optimizing with global CZ fidelity remains effective, leading to significant improvements in CZ $A$, CZ $C$, global fidelity, and correlation.
Considering single CZ gate fidelities, the fidelity of CZ C improves from 93.78\% to 96.19\%. The gate error is relatively reduced by 38.75\%. Regarding the two CZ gate correlations, the $correlation_{AC}$ reduces a lot from 1.15\% to 0.78\%, relatively varying 32.17\%. The results show an advantage when optimizing with global fidelity.

\begin{table}
\centering
\caption{Fidelity and correlation data of 3-pair parallel CZ gates in four different classes. `Global' and `Local' denote using global fidelity and local fidelities for optimization, respectively; `ref' and `iter' denote reference fidelity and iterative fidelity, respectively. The mean and standard deviation values are calculated based on the fidelity data during iterations 100-180.}
\begin{tabular}{ccccccccc}
\hline
  & \multicolumn{8}{c}{Fidelity} \\
  \cline{2-9} 
\multirow{2}*{\shortstack{Optimization\\ class}} & \multicolumn{2}{c}{CZ A} & \multicolumn{2}{c}{CZ B} & \multicolumn{2}{c}{CZ C} & \multicolumn{2}{c}{Global} \\
  \cline{2-9} 
~ & Mean & SD & Mean & SD & Mean & SD & Mean & SD \\
  \hline
Global ref & 94.93\% & 0.46\% & 95.52\% & 0.28\%  & 93.78\% & 0.63\% & 88.74\% & 0.70\% \\
Global iter & 96.21\% & 0.27\% & 95.94\% & 0.21\%  & 96.19\% & 0.19\%  & 92.04\% & 0.30\% \\
Local ref & 94.96\% & 0.40\% & 95.42\% & 0.36\%  & 93.83\% & 0.73\% & 88.79\% & 0.98\% \\
Local iter & 94.96\% & 0.40\% & 94.15\% & 0.70\%  & 94.07\% & 0.71\%  & 87.65\% & 1.20\% \\\hline
\end{tabular}

\begin{tabular}{ccccccccc}
\hline
  & \multicolumn{8}{c}{Correlation} \\
  \cline{2-9} 
\multirow{2}*{\shortstack{Optimization\\ class}} & \multicolumn{2}{c}{AB} & \multicolumn{2}{c}{AC} & \multicolumn{2}{c}{BC} & \multicolumn{2}{c}{ABC} \\
  \cline{2-9} 
~ & Mean & SD & Mean & SD & Mean & SD & Mean & SD \\
  \hline
Global ref & 1.26\% & 0.17\% & 1.15\% & 0.21\% & 1.45\% & 0.16\% & 3.69\% & 0.31\% \\
Global iter & 1.06\% & 0.10\% & 0.78\% & 0.10\% & 1.48\% & 0.11\% & 3.22\% & 0.16\% \\
Local ref & 1.28\% & 0.19\% & 1.14\% & 0.20\% & 1.50\% & 0.16\% & 3.79\% & 0.31\% \\
Local iter & 1.18\% & 0.24\% & 1.14\% & 0.20\% & 1.40\% & 0.25\% & 3.53\% & 0.39\% \\\hline
\end{tabular}
\label{tab:opt3CZ}
\end{table}

For the 2-pair parallel CZ gates, we also present the detailed fidelity and correlation data calculated from the results of 100 to 180 iterative steps in the optimization process. The results are shown in Table~\ref{tab:opt2CZ}. The average values of fidelity and correlation associated with the reference parameters are very close for optimizing with global fidelity and the individual local gate fidelities, indicating the experimental environment is almost the same for the two optimization processes. This is the precondition for us to compare the optimization results fairly. Focusing on the iterative fidelity, neither of the objective functions leads to a significant improvement on CZ I, i.e., the first CZ gate. Conversely, the results for CZ II, or the second CZ gate, show a fidelity improvement of $3.43\%$ and $3.22\%$ for optimizing with global fidelity and the individual local gate fidelities, respectively. This suggests that both objective functions effectively capture errors in individual gate parameters, and optimizing with global fidelity yields better results. This can also be seen from the correlation data. The absolute value and the standard deviation of the correlation decrease more when optimizing with global fidelity.

\begin{table}
\centering
\caption{Fidelity and correlation data of 2-pair parallel CZ gates in four different classes. The meaning of the classes is the same as that in Table~\ref{tab:opt3CZ}. The mean and standard deviation values are calculated based on the fidelity data during iterations 100-180.}
\begin{tabular}{ccccccccc}
\hline
~ & \multicolumn{6}{c}{Fidelity} & \multicolumn{2}{c}{Correlation} \\
  \cline{2-9} 
\multirow{2}*{\shortstack{Optimization\\ class}} & \multicolumn{2}{c}{CZ I} & \multicolumn{2}{c}{CZ II} & \multicolumn{2}{c}{Global} & \multirow{2}*{Mean} & \multirow{2}*{SD} \\
\cline{2-7}
~ & Mean & SD & Mean & SD & Mean & SD & ~ & ~ \\\hline
Global ref & $95.90\%$ & 0.29\% & $92.19\%$ & 1.21\% & $88.50\%$ & 1.42\% & $0.08\%$ & 0.61\%\\
Global iter & $95.98\%$ & 0.32\% & $95.62\%$ & 0.46\% & $91.80\%$ & 0.69\% & $0.02\%$& 0.38\% \\
Local ref & $96.03\%$ & 0.32\% & $92.10\%$ & 1.34\% & $88.35\%$ & 1.59\% & $-0.09\%$ & 0.69\% \\
Local iter & $95.88\%$ & 0.41\% & $95.32\%$ & 1.55\%  & $91.09\%$ & 1.61\% & $-0.30\%$ & 0.62\% \\\hline
\end{tabular}
\label{tab:opt2CZ}
\end{table}

%%%%%%%%%%%%%%%%%%%%%%%%%%%%%%%%%%%%%%%%
\bibliographystyle{apsrev}
%%%%%%%%%%%%%%%%%%%%%%%%%%%%%%%%%%%%%%%%

%%%%%%%%%%%%%%%%%%%%%%%%%%%%%%%%%%%%%%%%
% \bibliography{bibPara.bib}
%%%%%%%%%%%%%%%%%%%%%%%%%%%%%%%%%%%%%%%%